\documentclass[12pt]{amsart}
\usepackage{fullpage}
\usepackage{float,rotating}
\usepackage{caption}
\usepackage{parskip}
\usepackage{xcolor}
\usepackage{subcaption}
\usepackage{breqn}
\newenvironment{tablenotes}[1][Note]{\begin{minipage}[t]{\linewidth}\footnotesize{\itshape#1: }}{\end{minipage}}
\PassOptionsToPackage{hyphens}{url}
\usepackage[colorlinks = true,
            linkcolor = blue,
            urlcolor  = blue,
            citecolor = blue,
            anchorcolor = blue]{hyperref}

\usepackage{graphicx}
\usepackage{longtable}
\usepackage{booktabs,array,multirow}

\usepackage[noamsmath]{kpfonts}
\usepackage{inconsolata}
\usepackage[T1]{fontenc}

\usepackage{amsfonts,amsmath,amssymb}
\usepackage{amsbsy}
\usepackage{amstext}
\usepackage{amsthm}
\usepackage{nicefrac}
\usepackage{enumitem}

\newtheorem{hypoth}{Hypothesis}
\newtheorem{result}{Result}

\newtheorem{definition}{Definition}
\newif\iflatexml\latexmlfalse
\AtBeginDocument{\DeclareGraphicsExtensions{.pdf,.PDF,.eps,.EPS,.png,.PNG,.tif,.TIF,.jpg,.JPG,.jpeg,.JPEG}}
\usepackage{url}
\mathchardef\mhyphen="2D
\usepackage[authoryear]{natbib}
\usepackage{array}

\newcolumntype{P}[1]{>{\raggedleft\arraybackslash}p{#1}}
\newcolumntype{C}[1]{>{\centering\arraybackslash}p{#1}}

\newcommand{\SD}[1]{{\tiny $\left(#1\right)$}}

\newcommand{\mEst}[2]{\underset{\mbox{\SD{#2}}}{#1}}

\newcommand{\G}[1]{\ifthenelse{#1=1}{PD1}{\ifthenelse{#1=2}{PD2}{\ifthenelse{#1=3}{$\Sigma$-DOM1}{$\Sigma$-DOM2}}}}
\newcommand{\GR}[1]{\ifthenelse{#1=5}{PD3}{PD4}}

\def\ShowContent{1}
\newcommand{\ShowIf}[1]{\if\ShowContent1 #1 \fi}

\usepackage{setspace}

\begin{document}
\ShowIf{
\title{The Experimenters' Dilemma: Inferential Preferences over Populations}
\author{Neeraja Gupta}
\author{Luca Rigotti}
\author{Alistair Wilson}
\thanks{We would like to thank: David Huffman, Sera Linardi, and Lise Vesterlund.}
\date{July 2021}

\maketitle

\begin{abstract}
        We compare three populations commonly used in experiments by economists and other social scientists: undergraduate students at a physical location (lab), Amazon's Mechanical Turk (MTurk), and Prolific. The comparison is made along three dimensions: the noise in the data due to inattention, the cost per observation, and the elasticity of response. We draw samples from each population, examining decisions in four one-shot games with varying tensions between the individual and socially efficient choices. When there is no tension, where individual and pro-social incentives coincide, noisy behavior accounts for 60\% of the observations on MTurk, 19\% on Prolific, and 14\% for the lab. Taking costs into account, if noisy data is the only concern Prolific dominates from an inferential power point of view, combining relatively low noise with a cost per observation one fifth of the lab's. However, because the lab population is more sensitive to treatment, across our main PD game comparison the lab still outperforms both Prolific and MTurk.
\end{abstract}
}
\section{Introduction}
Experiments have been used extensively to uncover many facets of economic decision-making that would be veiled in naturally occurring data. Over the previous half-century, the dominant paradigm was the laboratory experiment: a set of participants (often undergraduates) recruited to a fixed time-slot and physical location, where a tailored set of monetary incentives are then offered to examine and identify economic hypotheses. In the last decade though a number of online populations have emerged for conducting economic experiments on. These populations offer an array of positives: greater convenience, lower barriers to entry,  more representative populations, etc. Moreover, they also offer lower costs per observation than lab studies. However, one concern is that reduced control might lead to less useful data.

While good experimental design requires a researcher to distill the environment down to its brightest, clearest essence, near any study that seeks to generate an incentivized decision requires  active engagement/understanding from the participants.  In the lab, distractions and outside-options activities can be minimized; the pace of the instructions can be set to maximize understanding, etc. A laboratory study can corral the participants attention but at a price: the greater time commitments have to be paid for with higher per participant payments. 

Attaining this level of control on online platforms is often thornier. However, as a trade-off, typical expected payments per participant on online platforms like Amazon's Mechanical Turk (MTurk) are often much smaller, reflecting the shorter time commitments. This creates a double edge. On the one-hand, each study provides small monetary incentives, with participants driven more by the volume of studies completed than any single one of them. If participants reduce how much attention they devote to any particular study, this attenuates the size of any treatment effect, and thereby hurts the very inference the study strives for. On the other hand, lower costs per observation allow a researcher to collect more observations for the same fixed hit to their research budget, strengthening power. In the end, this raises a relatively well-defined choice problem from the researchers point of view: in terms of inferential power, where is the analyst best-off when spending their limited budget?

This paper compares the answers to the same economic question across three populations. Our first population is a standard laboratory study, with undergraduate students recruited to come to a physical lab. Our second two populations use online participants:  MTurk, likely the most ubiquitous online labor populations; and Prolific, an emerging platform with a more curated set of participants. Using a set of simple two-player strategic choices we compare the three populations over noise in decision-making----considering noise as the proportion of decision-makers that act independently of incentives. We measure this through behavior in two games where participant lack any effective strategic tension: self-interested behavior is entirely coincident with pro-social behavior. Taking as given that participants have underlying preferences that are increasing in both their own and the total payoffs, decisions against both the self-interest \emph{and} efficiency identify inattentive participants. Moreover, a separate source of variation in our experiments manipulates the first-listed option, thus creating a frame orthogonal to incentives. As such, we can decompose the extent to which participants in each population choose randomly across decisions, or simply choose the first-listed option available to them. 

While two of our games provide a diagnostic over inattention, the second two games embed a hypothesis with extensive prior evidence. We use a comparison of behavior over these two games to answer our final question: given a fixed budget, which  population is preferred by a researcher interested in making a qualitative directional inference? Here we use a canonical trade-off between efficiency and own-payoff, with both games having a prisoner's dilemma (PD) structure. Reflecting a literature that emphasizes partial cooperation, engaging cooperatively so long as the trade-offs between own-payoff and social efficiency are not too large, our chosen games vary the relative temptation to defect. We use the prior literature to form expectations over the size of the treatment effect, and thus induce the analyst's preference over populations via the inferential power in a test of the differences in play across the two games. In particular, we parameterize each experimental population with two properties: (i) the cost per independent observation in dollars; and (ii) the size of the attenuation over the treatment effect. We then form the analyst's preference as if setting up a standard choice problem: drawing  both iso-power contours under a fixed budget (analogous to the indirect utility) and the dual iso-budget contours under a fixed power level (analogous to the expenditure function).

Fixing the expected cost while creating standard incentives for each population (via a shift in the probability of payment) we then examine which of the populations is preferred. While our overall results outline different strengths for both Prolific and the Laboratory, they suggest that MTurk is dominated by Prolific. Our estimates indicate that 60 percent of the MTurk participants make decisions that are entirely independent of the induced incentive (either through random choice, or choosing the first-available choice). This compares with estimates of 14 percent in the laboratory and 19 percent on Prolific (neither of which shows any bias towards first-listed options).

Despite the large amount of noise on Mturk, ceteris paribus, this population should still have greater inferential power than the laboratory. This reasoning is driven by differential costs: an average observation costs \$3, \$4.40 and \$22 on MTurk, Prolific and the lab, respectively. Holding constant the experiment's budget, the inferential power from a low-noise sample of 75 laboratory participants must be compared to the inferential power from a high-noise sample of 550 participants on MTurk (or 380 low-noise participants on Prolific). Were attenuation in treatment-effects purely driven by the inattentive proportion then Prolific is the clearly dominant population, followed by MTurk, and then the lab sample as the least effective. For a 90 percent power test (with 90 percent confidence on the two-sided comparison) the necessary experimental budget to generate a significant effect for our PD comparison would be \$660 on Prolific, \$1,830 on MTurk and \$2,940 in the lab. 

But the proportion of inattentive/noisy participants is not the entire story. The two online populations also exhibit reduced elasticity of response: a smaller effect-size under the same induced treatment. Factoring in this reduced quantitative size of the response, our two online populations end up having diminished inferential power relative to the laboratory.  Despite the low cost per observation on Prolific and relatively high attentiveness, the sample is just too inelastic. Both online samples exhibit fairly extreme other-regarding behavior, with a near negligible response to a shift in the PD tensions. What the literature led us to think of as a moderately sized exogenous treatment for the lab ends up being far too subtle for the online samples. As a final robustness check, to make sure it is inelasticity rather than a true null, we induce a much starker PD treatment on Prolific in an supplemental experiment. While we do find a significant effect on Prolific under the more sweeping treatment comparison, the elasticity of response continues to be quantitatively much smaller than comparable lab studies. However, despite the smaller treatment effect, the difference is now large enough that the small cost per observation in the Prolific sample dominates the laboratory in power terms (just).

Some of our results are likely specific to social dilemmas, and the potentially higher level of attention required in strategic settings. However, the environments we examine are simple enough that the very noisy data from MTurk points to that population as being dominated by more-curated online populations such as Prolific. Despite Prolific being almost 50 percent more expensive per observation, the additional signal more than compensates for this. Particularly for very stark economic comparisons, the ability to collect many, many observations favors online populations such as Prolific. However, our study also points to a potential benefit of laboratory samples. Despite the expense, lab observations seem to exhibit much-greater sensitivity to changes in the offered incentives. For more-nuanced hypotheses or more-complicated environments, lab samples may well be preferable.

The paper is organized as follows: Section \ref{sec:lit} summarizes the related literature and highlights our contributions, section \ref{sec:design} discusses our experiment design, main hypothesis, incentives and implementation for all three samples. Section \ref{sec:results} discusses our main results and section \ref{sec:conc} concludes.

\section{Related Literature}\label{sec:lit}
A number of studies have compared MTurk and laboratory population, primarily focused on whether empirical regularities observe in the lab can be replicated. Our paper novelty is in both adding another population (Prolific) but also in making the focus more explicitly on the effective power on each population, taking into account researchers' financial constraints. In particular, we consider how the possibilities for many more independent observations from cheaper online populations interaction with the potential for noisier data, or a more inelastic response.

One of the earliest works examining the use of MTurk in online experiments is \cite{paolacci2010running}, replicating three classical behavioral economics results (the Asian disease, Linda and Physician problems), and finding no significant differences between the populations. Along similar lines, \cite{horton2011online} find no significant differences in cooperation between an MTurk sample and the experimental lab literature on one-shot PD games. In \cite{goodman2013data} an MTurk sample replicates standard decision‐making biases.\footnote{MTurk participants exhibit: (i) present bias; (ii) risk‐aversion for gains and risk‐seeking for losses; (iii) show delay/expedite asymmetries; and (iv) show the certainty effect.} More recently, \cite{thomas2017validity} suggests that strict exclusion criterion for ``problematic'' participants can reduce statistical noise without introducing sampling bias. In \cite{arechar2018conducting} the researchers uncover the same basic behavioral patterns of cooperation and punishment in a repeated public good experiment in both the lab and MTurk, even though dropout can be a challenge for conducting interactive experiments on MTurk. 

\cite{snowberg2018testing} elicit and compare a battery of behavioral attributes using a survey administered to an entire undergraduate cohort, a self-selected lab sample, and a representative sample of US online participants recruited from MTurk. While they look at many different behaviors, their elicitation dpes includes two one-shot PD games (though  with the same effective incentives). Similar to our findings, they do find significant differences in cooperation levels across populations for the PD game, with the online sample being more cooperative, however they do find comparable comparative statics across populations across many other behaviors. Their other overarching results are that behavioral characteristics are similarly correlated across populations and that noise (as measured by difference in response for duplicate elicitations) is higher for online populations. We confirm this last result when it comes to MTurk but not for Prolific, though where our own measures of noise are based on responses to a more-basic check of rationality and a response to a frame change. Our focus is also switched, where we take as given that the effect, and instead focus on the effective power of the population under a fixed researcher budget. 

Our findings match growing concerns over a potential decline in the quality of MTurk data over the past two years. The literature highlights limitations of MTurk including, but not limited to, anticipation of deception by researchers, repeated participation in similar tasks leading to knowledge acquisition and a resultant change in behavior, unmeasurable attrition and programmed bots \citep{hauser2019common,chmielewski2020mturk}. \cite{aruguete2019serious} identify MTurk workers as being more likely to fail attention checks designed to measure haste and carelessness in responses than college students (though our measures of noise are on a sample that has successfully passed an understanding quiz). Some of our results echo this, and make this more concrete through our focus on inferential power. While we do find low inferential power for online data for both MTurk and Prolific, our noise measures point to Prolific as being close to the laboratory levels. Instead, the low-power on Prolific seems to come about through an inelastic response, where the population in general offers much more promise.
\section{Experiment Design}\label{sec:design}
\ShowIf{
\begin{table}[tb]
    \centering
    \caption{Experiment Design}
    \label{tab:Subjects}
    \begin{tabular}{ccccc}\toprule
   \textit{Panel A} &  \multicolumn{4}{c}{Payoff $\pi_i$ on action $(a_i,a_j)$}    \\\cmidrule{2-5}
        & $(C,C)$ & $(C,D)$ &  $(D,C)$ & $(D,D)$ \\ \midrule
    Game \G{1} & \$21 & \$2  & \$28 & \$8 \\
    Game \G{2} & \$19 & \$8  & \$22 & \$9 \\
    Game \G{3} & \$17 & \$12 & \$16 & \$10 \\
    Game \G{4} & \$15 & \$16 & \$10 & \$11\\
    \midrule
\textit{Panel B} & & \multicolumn{3}{c}{ Participants \& Expenditure}  \\\cmidrule{3-5}
         & & Lab & MTurk & Prolific \\ \midrule
    \multicolumn{2}{l}{\textbf{Participants:} }\\
        \multicolumn{2}{l}{\qquad $C$-first frame}    & 50   & 368 & 250\\
       \multicolumn{2}{l}{\qquad $D$-first frame} & 24  & 180 & 135\\ \cmidrule{3-5}
 \multicolumn{2}{l}{\qquad Total} & 74 & 548 & 385 \\ \midrule
 \multicolumn{2}{l}{\textbf{Expenditure:} }\\
       \multicolumn{2}{l}{\qquad Total} & \$1,634 & \$1,649 & \$1,680\\
        \multicolumn{2}{l}{\qquad Per observation} & \$22.08 & \$3.01 & \$4.36 \\
        \bottomrule
    \end{tabular}
    \begin{tablenotes}
Participant numbers exclude those who failed to answer comprehension questions correctly, though fixed-payment costs for these hits are included in the expenditure. \end{tablenotes}
\end{table}
}
Our experiment has core $3\times 2$ design over:
\begin{description}
    \item[Population] We use three populations: (i) students recruited from the University of Pittsburgh's undergraduate population  (the Lab sample); (ii) online workers recruited from Amazon's online marketplace Mechanical Turk (the MTurk sample); and (iii) online workers recruited from Prolific (the Prolific sample).
    \item[Strategic environment] We ask participants to make a binary choice across four symmetric two-player games (with payoffs provided in Table \ref{tab:Subjects}). While our experiment uses an $A$/$B$ action labeling, we use a $C$(ooperate)/$D$(efect) labeling in the paper as all four games have joint cooperation as the socially efficient outcome.\footnote{Games are presented to participants in a random order.}
    \item[Irrelevant Frame] The frame variable changes the order in which the actions are presented to subjects, permuting the ordering of the $C$ and $D$ actions in the presented game tables.\footnote{We present games to subjects as a table with four rows (one for each possible action profiles) ordered as $AA$, $AB$, $BA$, $BB$ for the self/other action. The re-framing therefore moves the socially efficient $CC$ entry from the top entry in the table (labeled $AA$ in the experiment) to the bottom entry (labeled $BB$).} 
\end{description}

In terms of our hypotheses, we first outline the features of the games subjects play. All four games are dominance solvable in terms of individual payoffs (relabeling the standard notion of strict dominance):
\begin{definition}[$i$-Dominated action]
    Action $a$ is $i$-dominated if there exists another action $a^\prime$ that gives player $i$ higher payoff for any action of the other players.
\end{definition}
The $i$-dominant action profile (also the Nash action profile) is to defect in games \G{1} and \G{2} and cooperate in \G{3} and \G{4}. However, there is a large body of evidence suggesting that many individuals' preferences are other-regarding and sensitive to social efficiency. This evidence shows that many individuals choose $i$-dominated actions if these choices improve efficiency (as measured by the sum of payoffs). As such, a stronger version of dominance can be based on individual and total payoffs: 
\begin{definition}[$\Sigma$-Dominated action]
    Action $a$ is $\Sigma$-dominated if there exist another action $a^\prime$ such that $a$ is $i$-dominated by $a^\prime$, and the sum of the players' payoffs is smaller for $a$ than for $a^\prime$ for any action of the other players.
\end{definition}
Games \G{3} and \G{4} are constructed so that the $D$ action is $\Sigma$-dominated, and this action choice is thus hard to justify with almost any other-regarding preference.\footnote{Game \G{3} is designed to satisfy an even stronger ordering over unilateral deviations: the Pareto order. However, we do not find that this has any additional predictive content, so we focus purely on $\Sigma$-Dominance.}    Taking as given that participants on all populations are driven by a preference that is strictly increasing in both the own and social payoff, the only justification for $\Sigma$-dominated behavior is therefore that the agent did not fully understand the environment.

These two games therefore provide our first measure of attentiveness in each population, where our null hypothesis is therefore that:
\begin{hypoth}[Dominated-play null]
The three populations have similarly small proportions of $\Sigma$-dominated play.
\end{hypoth}
Our second measure for the quality of choices across the three populations is based on the response to the framing variable. A change in the order in which actions are presented changes nothing with respect to offered incentives. One plausible bias for inattentive participants is that they will move as quickly as possible through the offered choices by selecting the first-available option. A comparison of cooperation rates in the $\Sigma$-DOM game pair across the frame change therefore identifies this feature, where we would expect greater cooperation in our main treatments where $C$ is the first listed option.

\begin{hypoth}[Reframing null]
The three populations have the same cooperation rates across the re-framing (but within population).
\end{hypoth}

Our first two hypotheses are about assessing the degree of inattentiveness, using choices where we can say that one option---separate from preferences which may legitimately vary across populations option---is a priori dominated. Our final hypothesis is more nuanced, and instead concerns the inferential power over a relatively standard hypothesis (albeit where the evidence tends to come from lab studies). In this hypothesis we aim to compare the behavior between games \G{1} and \G{2}, where we change the intensity of the prisoner's dilemma tensions. 

To form our hypothesis we make use of a parametric index known as the Rapoport ratio \citep[cf.][]{rapoport1967note}, which has been shown to be predictive of cooperation. The Rapoport ratio is given by a function of the PD-game payoffs: $$\rho=\tfrac{\pi_i(C,C)-\pi_i(D,D)}{\pi_i(D,C)-\pi_i(C,D)}.$$ 
The behavioral literature indicates that the frequency of cooperation is increasing with this ratio, where games \G{1} and \G{2} have Rapoport ratio's of 0.50 and 0.71, respectively. As such we would expect cooperation rates to be greater in \G{2} than \G{1},  and the aim of inference will be to identify a significant \emph{directional} effect. However, our comparison across populations is focused on the differential power of this test.

To assess power, we need a baseline assumption on the true quantitative size of the effect. To do this, we make use of the one-shot PD game results reported in \citet{charness2016social} from laboratory data. From their data we would expect to see a cooperation increase of 14 percentage points as we move from  game \G{1} to \G{2}.\footnote{We estimate this from \citeauthor{charness2016social} via a logit model with the Rapoport ratio as the sole predictor. The estimated model predicts a cooperation rate given by: 
$$\text{Coop}(\rho)= \frac{1}{1+5.66\cdot e^{-3.32\rho}}.$$
} Across our three populations we can therefore specify the directional comparative-static that such an experiment might set out to uncover within each population:

\begin{hypoth}[PD comparative static]\label{hyp:Rapoport}
Following the Rapoport ratio prediction, each of the three populations will have more cooperation in game \G{2} than \G{1}.
\end{hypoth}

While this hypothesis is intuitive and supported by prior evidence, it can also be used to generate a horse race across the populations, by examining their inferential power. Given that decisions are binary, tests over the difference across \G{1} and \G{2} are simply a function of the observed cooperation rates in each game and the sample size $N$.\footnote{Greater statistical power can be generated if we also use the within-subject nature of the data, however, for simplicity we focus on a more-standard between comparison} The $T$-statistic for a test of a null effect between the two games for two samples of size $N$ with cooperation rates of $P_1$ and $P_2$ is given by:
\begin{equation}\label{formula:Tstat}
    T\left(P_1,P_2,N\right)=\frac{\sqrt{N}\cdot\left(P_2-P_1\right)}{\sqrt{
\left(P_1+P_2\right)\left(1-\tfrac{P_1+P_2}{2}\right)}}.
\end{equation}
For a qualitative alternative hypothesis that there is more cooperation in game \G{2},  we would therefore want the $T$-statistic to be greater than approximately 1.64 to attain significance at 95 percent confidence (or 90 percent on the two-sided alternative).

Modeling the number of cooperation decisions within the sample $N\cdot P_1$ and $N\cdot P_2$, as binomial draws with true proportions $p_1$ and $p_2$, respectively, it is therefore possible to calculate the likelihood that we would make a type-II error on this $T$-test when the two populations are in fact different. Using the \citeauthor{charness2016social} data, if the true cooperation rates for games \G{1} and \G{2} are 0.48 and 0.65, respectively, then the power of the test is a direct function of the sample-size $N$. Given a fixed experimental budget, all else being equal, whichever population has the cheapest observations would yield the greatest power.

But herein lies the rub. Cheaper data is plausibly noisier. Certainly, there is a general wariness of online samples, with the thought that reduced control---for example,  the ability for participants to multi-task while taking part in the study---leads to a larger proportion of participants being inattentive, or unresponsive to the economic treatment. To model this, we consider a population as having two fundamental properties: a dollar cost per observation $c$; and a noise/attenuation parameter $\gamma$. The cost per observation is from the point of view of the analyst who we will assume has a finite budget, while the attenuation parameter $\gamma$ affects the population-level expectation in PD game $j$, attenuating it towards a coin flip via  $\gamma\cdot\tfrac{1}{2}+(1-\gamma)\cdot p_j$.\footnote{The parameter $\gamma$ here represents any form of attenuation. While one obvious source of attenuation here is inattention to the experimental environment, the parameter can also be interpreted as reduction in the elasticity of response within the population, as this will also reduce the quantitative size of the treatment response.}

Our assessment across the populations will therefore take into account their costs, and will focus on the inferential power. We will therefore think of the analyst's problem through the lens of a consumer-choice-like problem, with statistical power in place of utility. Population $A$ is preferred to population $B$ if it provides greater statistical power under a fixed research budget. Populations are characterized by a cost/noise pair $(c,\gamma)$, leading to a well-defined probability of making a Type-II error on the $T$-stat in \eqref{formula:Tstat}.\footnote{We calculate the affordable sample size $N$ from the total budget, while the sample proportions $P_1$ and $P_2$ are both attenuated towards $\nicefrac{1}{2}$ at the rate $\gamma$}. Using the idea that the analyst's preferences are represented by statistical power (the probability of \emph{not} making a type-II error) in Figure \ref{fig:PowerTheory} we indicate indifference curves in $(c,\gamma)$-space for Hypothesis \ref{hyp:Rapoport}. In particular in panel (A) we indicate iso-power contours under a fixed experimental budget (\$1,650, the approximate budget in our experiments), analogous to thinking about the indirect utility function in consumer choice. In panel (B) we indicate iso-budget lines under a fixed power level (90 percent), analogous to the expenditure function in the dual consumer choice problem. 
\ShowIf{
\begin{figure}[tb]
    \centering
     \begin{subfigure}{0.49\textwidth}
        \includegraphics[width=1\textwidth]{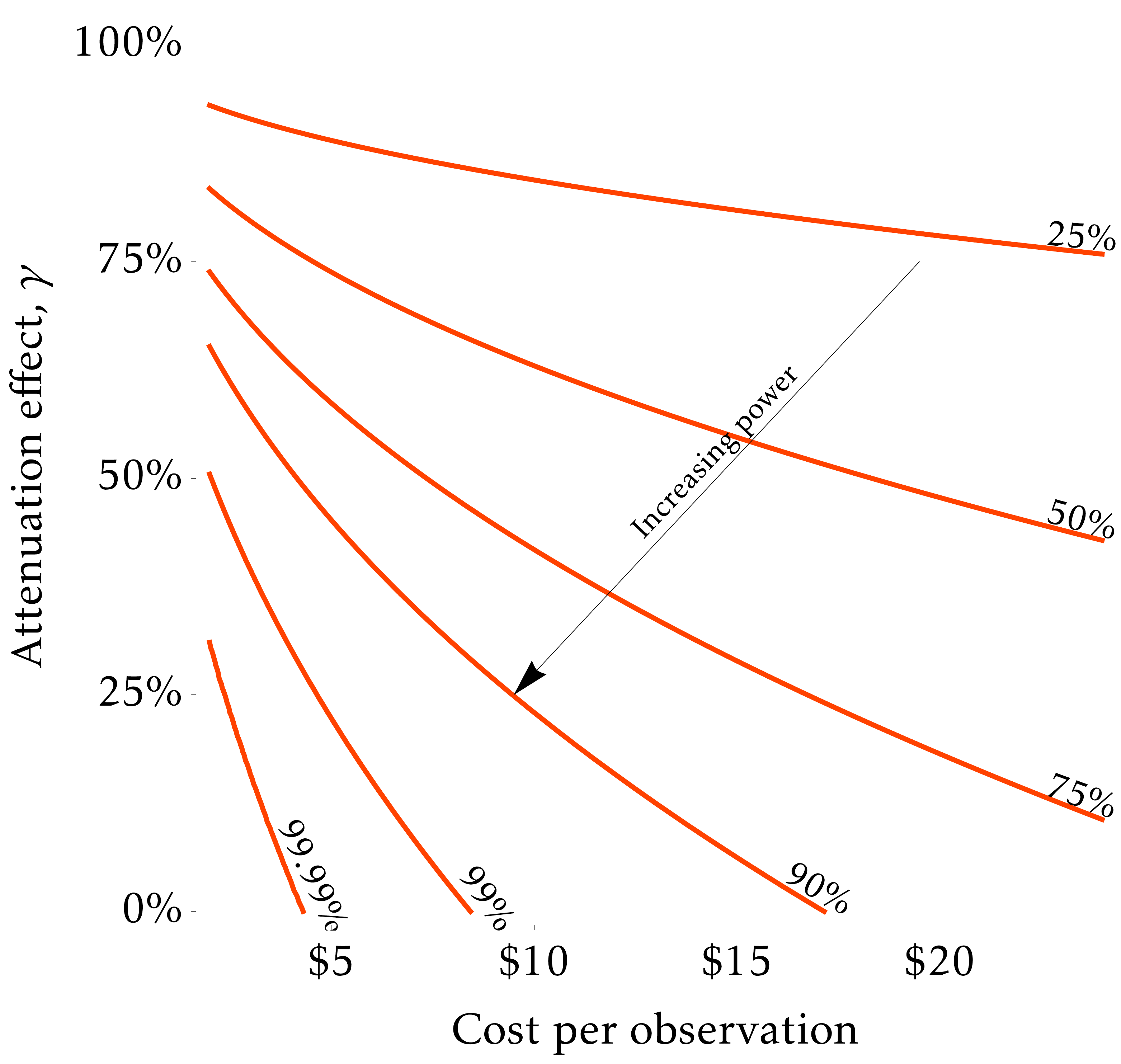}
        \caption{Iso-Power Contours (\$1,650 budget)}
    \end{subfigure}
     \begin{subfigure}{0.49\textwidth}
        \includegraphics[width=1\textwidth]{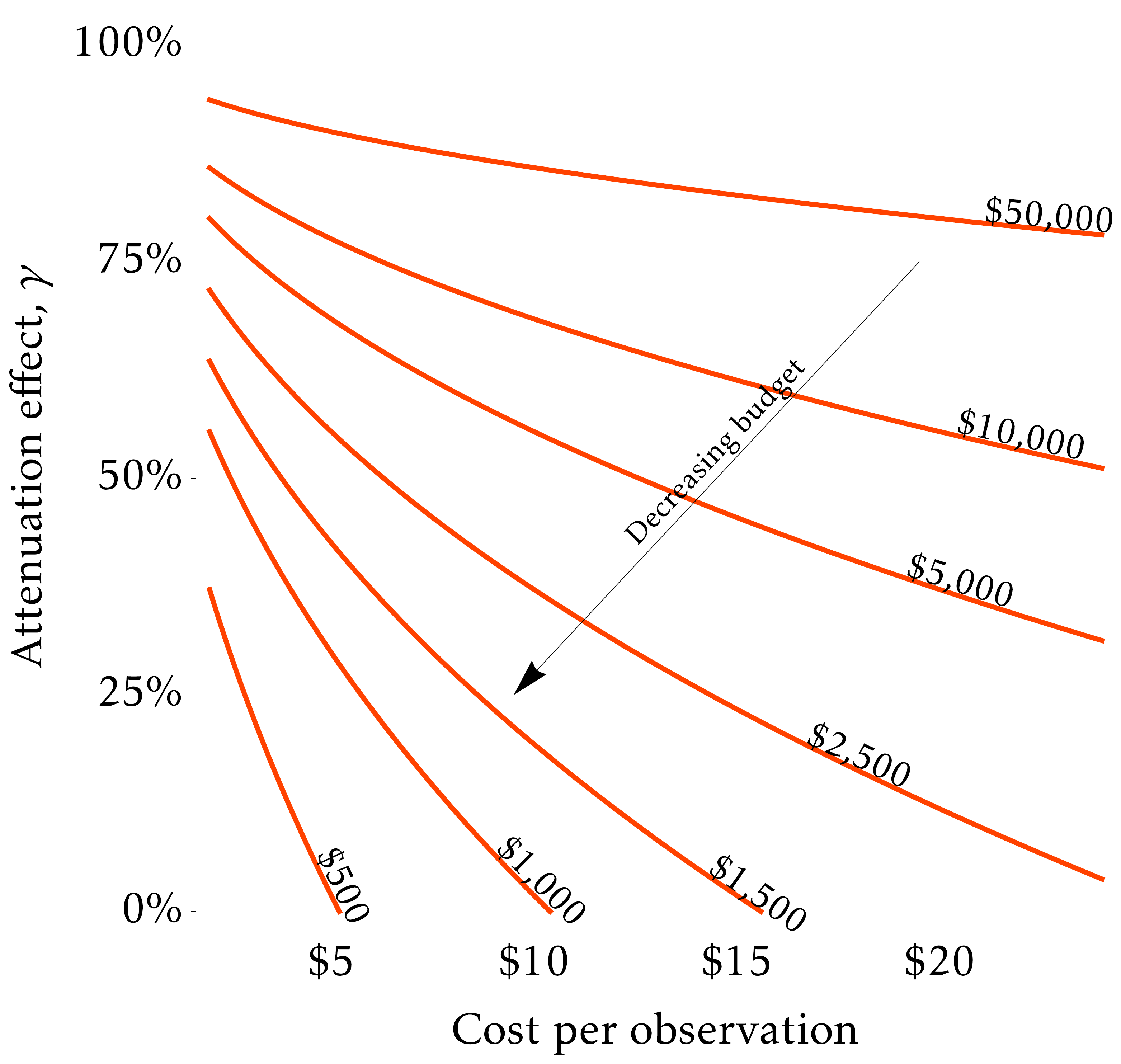}
        \caption{Iso-Budget Contours (90\% power)}
    \end{subfigure}    
        \caption{Analyst inferential preferences: Noise versus Cost\label{fig:PowerTheory}}
\begin{tablenotes}
Panel (A) shows iso-power contours (where labels indicate the probability of rejecting null) for an experiment with a \$1,650 budget, while panel (B) shows iso-budget contours for a test with 90 percent power using a two-sample $t$-test on the PD cooperation rates with population variables derived from \citet{charness2016social}.
\end{tablenotes}
\end{figure}
}

Both $\gamma$ and $c$ are `bads' from the analysts point of view, and satisfy local non-satiation. As such, better populations have the smallest possible values for each, and  the analyst's preference is increasing as we move from the top-right to the bottom-left  of Figure \ref{fig:PowerTheory}. The figure helps motivate the final piece of our design. Rather than fixing the sample size $N$ within each population we instead parameterize our experiments so that the per participant payments match standard rates on each population, but where we fixed the overall budget. This design choice, alongside the the identification of the population noise rates, allows us to run an inferential horse race over our populations. 

Essentially, we ask which populations lies on the researcher-best contour in Figure \ref{fig:PowerTheory}. For example, a sample with a cost per participant of \$17.50 and no noise has 90 percent power for our comparison of the PD games (the labels on each contour provide the type-II error probability). However, inspecting the figure, an online sample with a \$3 cost per observation is preferable even when half of the sample is pure noise as this sample would have 99 percent power under the same total expenditure. Below we detail the incentives offered to each population, and how budgets were determined.

\subsection{Incentives and Implementation} \label{sub:incentives}
As detailed above our design collects data across six between-subject treatments, the three populations and the frame-change over the ordering of the cooperate/defect decision. For our horse-race, our initial plans were for a budget of \$1,500 per population. However, our lab study was run first, and ended up being more expensive at just over \$1,600. We therefore matched the MTurk and Prolific samples to this approximate budget. Within this population budget, we then aimed to spend the money across the $C$-first/$D$-first frames at a two-to-one ratio (in case pooling the samples was not an option for the lab sample).\footnote{Focusing purely on the average earnings of the participants (so excluding fees and other costs) divided by the average time taken to complete each study, the effective wage rates are remarkably similar. Across the lab, MTurk and Prolific the effective wage rates are \$31.66, \$31.81 and \$40.25 per hour, respectively.}

Our lab sample was collected right before the pandemic hit, where we recruited University of Pittsburgh undergraduate students and ran in person. Participants were offered a \$6 fixed fee, and were randomly paid for one decisions over the four games after being matched to an anonymous partner.\footnote{The experiment was programmed in oTree (\cite{chen2016otree} and conducted at the Pittsburgh Experimental Economics Laboratory (PEEL).} Payments for the selected game were determined by Table \ref{tab:Subjects}. In total our expenditure for the 74 laboratory observations was \$1,624, where this figure includes \$72 spent on show-up fees for unused participants.\footnote{Our methodology here is to include all variable costs for a study incurred by the researcher. One possible critique here is that we do not account for the financial costs of setting up and running the PEEL lab, where our approach is to treat these as sunk costs. As such, inferential comparisons across populations are from the point of view of a researcher who has free access to a turn-key lab space.}

The per-observation cost for the Lab data are therefore just over \$22. While we could have offered these incentives to the MTurk participants, this would have represented a substantial break from the norm on this population. As our aim was to match the effective incentives being offered on each population and to account for this in inference, we scaled down the incentives on MTurk substantially. Participants in the MTurk sample were given a \$0.50 fixed fee and a further \$0.50 if they correctly answered two comprehension question to show that they understood the instructions.\footnote{Subjects who failed to answer the comprehension questions were not asked about game decisions and are excluded from out count of $N$ and analysis. However the \$0.50 costs for these subjects as well as the fees for Amazon (20\%) and Turk Prime \citep[4\%][]{litman2017turkprime} are included in the total expenditure. In terms of population, MTurk participants were restricted to a standard subsample: those located within the US, with a 95\% or better approval rate.}$^{,}$\footnote{The MTurk and Prolific samples were both run using Qualtrics.} While payments within each game table exactly matched the lab sample, as given in Table \ref{tab:Subjects}, payments were scaled down in the likelihood of payment. Pairs of participant were paid for their decisions from one of the four game tables with a 10 percent probability.\footnote{Our instructions gave participants a clear rule used to conduct randomizations, where all draws were made using public randomization outside of the researchers' control (here public state lottery draws made the evening after the decisions were taken). Moreover, because participants were told that if selected for payment they would be matched to another payment participant, where the final bonus-payment would be determined by the choices of both payment participants' choices. As such, conditional on payment, the externalities and lottery over the four games are identical to our lab study.} In total we collected data from 548 individuals with a total cost of \$1,649 (\$3 per participant).

Finally, our Prolific sample followed an identical process for the marginal incentives and game payment. However, rules for the platform required a larger minimum payment, so we increased the fixed payment to \$1.60.\footnote{Participants failing the comprehension check received the fixed payment, but were not given the marginal incentive. The total expenditure includes the costs for these participants as well as the 33\% fee imposed by the platform.}$^{,}$\footnote{We conducted a pilot of 20 participants on Prolific to understand the median time taken, where the minimum fixed-fee payment was a function of this time. However, the marginal incentives for this pilot were higher and the fixed fee lower). For the sake of comparability, neither this pilot data, nor its cost are considered in our analysis.} The total expenditure on Prolific was \$1,680 for 385 observations.

Our overall plan embeds an essential question: Given the differential costs for each observation, and the potential quality differences in the data collected, which population is superior? MTurk offers the potential for the largest number of observations given a fixed budget. However, it is also potentially the noisiest. On the other extreme, the laboratory is the most expensive per observation. The question is whether this additional expense is warranted through higher quality data.

\section{Results}\label{sec:results}
We now outline the three core results from the experiment, before presenting them in detail: (i) The laboratory and Prolific samples are similar over the fraction of participants making  $\Sigma$-dominated choices ($\sim$10 percent), while in the MTurk sample this proportion is much larger ($\sim$37 percent); (ii) Changing the order of actions has no significant effects in the lab and Prolific samples, while the MTurk sample exhibits a 16 percentage point swing in favor of the first-listed choice; (iii) The three samples detect a significant proportion of participants who choose a fully selfish $i$-dominant strategy profile in the four games; however, where the lab sample exhibits a standard response to increasing trade-offs between self and other, the MTurk and Prolific samples are essentially inelastic on this margin, with much stronger other-regarding behavior. We outline these results in turn below.

Table \ref{tab:PooledCoop} provides averages outcomes across the experiment (with standard errors derived from simple tests of proportion). In Panel A we first outline the proportion of individuals with particular focal behaviors over the four games (pooling data across the frame), then outline the relative effects across the re-framing. In Panel B we then repeat the exercise, but where we outline the average cooperation rates by game and population (again pooling data across the frame, and the change across the frame). Paralleling the two table panels, Figure \ref{fig:CoreChoice} provides an illustration of the choice profiles and game cooperation rates.

Our first core result is illustrated by Figure \ref{fig:CoreChoice}(A) (as well as the first row in Panel A of Table \ref{tab:PooledCoop}) the rate at which individuals in the experiment make an obvious mistake with respect to the offered incentives. While all of our games are dominance solvable over payoffs, a large literature documents other-regarding preference. As such, our first measure focuses on games \G{3} and \G{4} where there is no tension at all between payoffs to self and other.\footnote{Cooperation in game \G{3} satisfies an even stronger dominance notion, that the individual's unilateral choice can be ordering using the Pareto ranking. However, if anything we observe larger error rates in game \G{3} than \G{4}, we opt for the weaker criterion of $\Sigma$-dominance.}  The fraction of participants who make a $\Sigma$-dominated choice to defect in either (or both) of these games is therefore one measure of the data quality in each population.

\ShowIf{
\begin{figure}[tb]
    \centering
    \begin{subfigure}{0.49\textwidth}
        \includegraphics[width=1\textwidth]{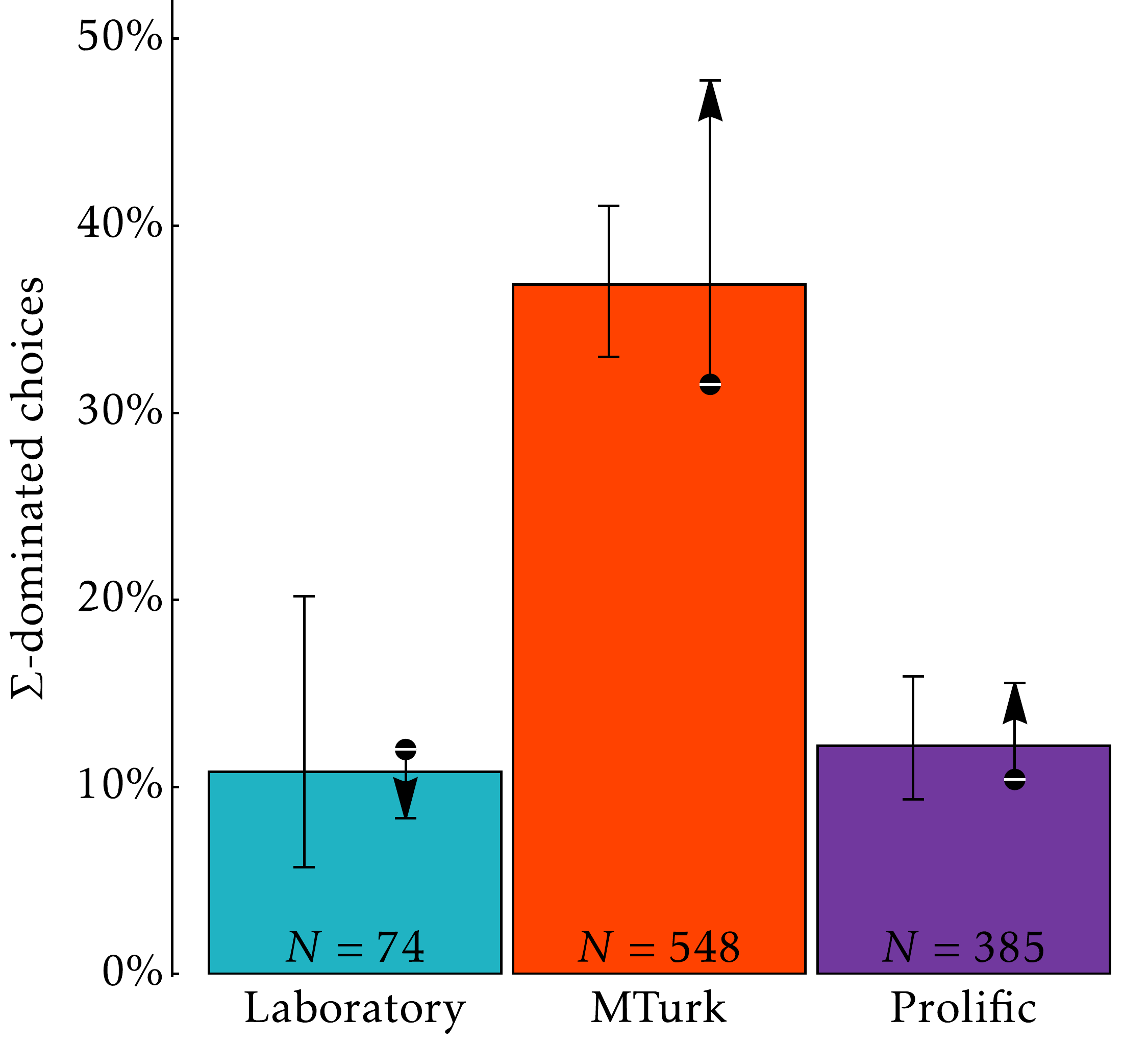}
        \caption{$\Sigma$-dominated individual choice}
    \end{subfigure}
    \begin{subfigure}{0.49\textwidth}
        \includegraphics[width=1\textwidth]{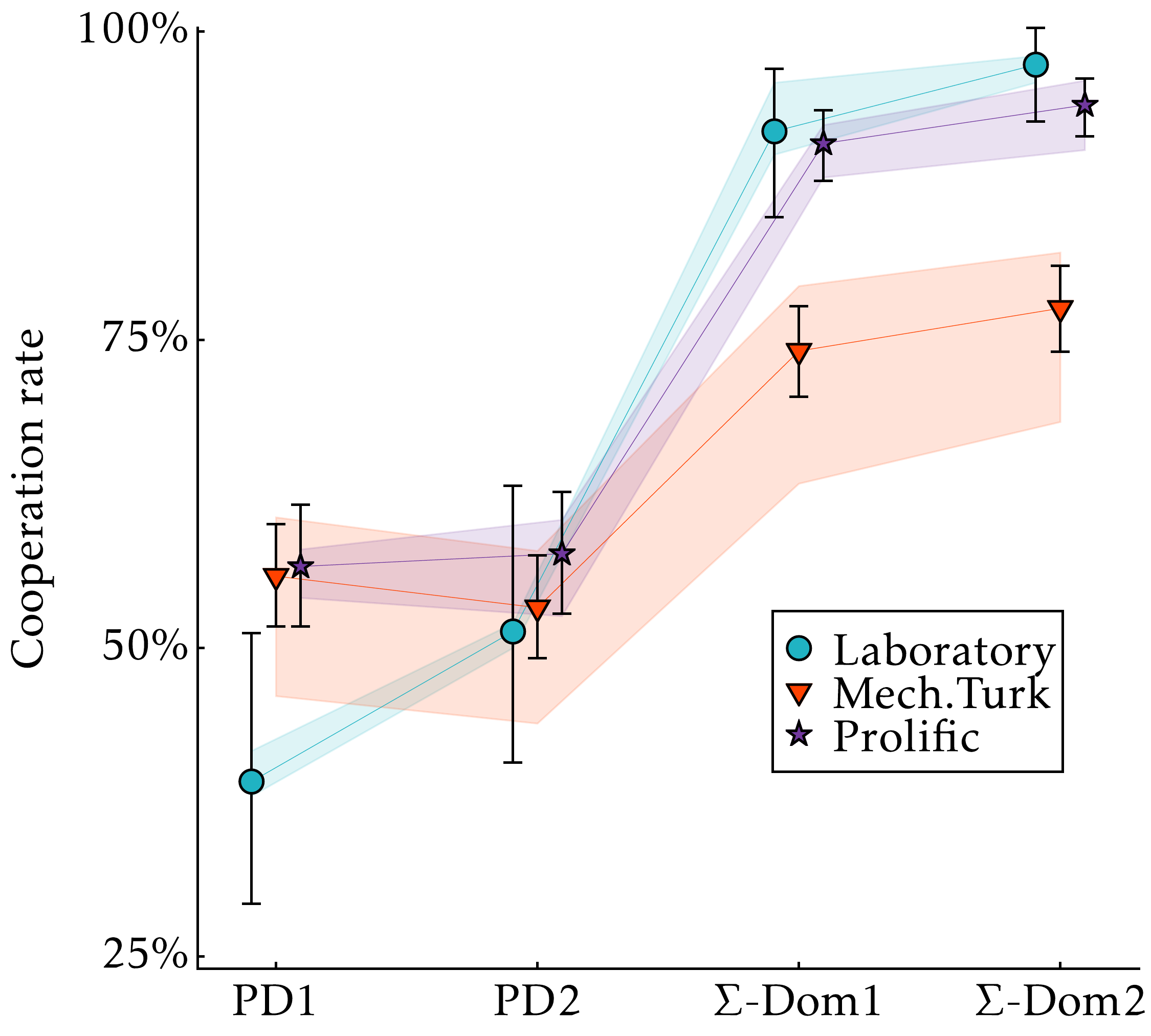}
        \caption{Cooperation rate by game}
    \end{subfigure}
    \caption{Average Behavior\label{fig:CoreChoice}}
\begin{tablenotes}
Error-bars indicate binomial-exact 95-percent confidence intervals for the proportion. Arrows in panel (A) and the shaded region in panel (B) indicate the change over the framing variable when the $C$ action is switched to be the listed after $D$ instead of before.
\end{tablenotes}
\end{figure}

\begin{table}[t]
\centering
\caption{Results Summary\label{tab:PooledCoop}}
\begin{tabular}{lcccccccc}
        \toprule & \multicolumn{2}{c}{Lab} & & \multicolumn{2}{c}{MTurk} & & \multicolumn{2}{c}{Prolific} \\\cmidrule{2-3}\cmidrule{5-6}\cmidrule{8-9}
                 & Avg & $\Delta_{\text{Frame}}$ & & Avg & $\Delta_{\text{Frame}}$ &  & Avg & $\Delta_{\text{Frame}} $ \\
        \midrule
        \textbf{$\Sigma$-Dominated:} & $\mEst{0.108}{0.036}$ & $\mEst{-0.037}{0.073}$  & & $\mEst{0.369}{0.021}$ & $\mEst{0.163}{0.044}$ & 
        &$\mEst{0.122}{0.017}$ & $\mEst{0.052}{0.037}$ \\ \cmidrule{2-9}
        \textbf{$i$-Dominant $(DD\mhyphen CC)$:} & $\mEst{0.324}{0.054}$ & $\mEst{0.075}{0.118}$ & & $\mEst{0.159}{0.016}$ & $\mEst{0.028}{0.034}$ & & $\mEst{0.260}{0.022}$ &  $\mEst{0.056}{0.048}$ \\
        \textbf{Rapoport identifier $(DC\mhyphen CC)$: }  & $\mEst{0.189}{0.046}$ &$\mEst{-0.033}{0.095}$ & & $\mEst{0.093}{0.012}$&$\mEst{-0.048}{0.024}$ & 
        & $\mEst{0.106}{0.016}$& $\mEst{0.056}{0.047}$ \\
         \textbf{Full Cooperator $(CC\mhyphen CC)$:} & $\mEst{0.284}{0.052}$& $\mEst{0.012}{0.112}$ && $\mEst{0.297}{0.020}$ & $\mEst{-0.046}{0.041}$ && $\mEst{0.416}{0.025}$ & $\mEst{-0.070}{0.052}$\\\cmidrule{2-9}
        \textbf{$\Sigma$-Dominant:} & $\mEst{0.892}{0.036}$ & $\mEst{0.037}{0.073}$& & $\mEst{0.631}{0.021}$ & $\mEst{-0.163}{0.044}$ & &$\mEst{0.878}{0.017}$ & $\mEst{-0.052}{0.037}$ \\
        \textbf{Rapoport ordered:} & $\mEst{0.905}{0.034}$ & $\mEst{0.017}{0.071}$& & $\mEst{0.828}{0.016}$ & $\mEst{0.049}{0.032}$ & &$\mEst{0.886}{0.016}$ & $\mEst{-0.018}{0.035}$ \\
        \textbf{Both:} & $\mEst{0.797}{0.047}$ & $\mEst{0.053}{0.096}$& & $\mEst{0.549}{0.021}$ & $\mEst{-0.066}{0.045}$ & &$\mEst{0.782}{0.021}$ & $\mEst{-0.040}{0.045}$ \\
        \bottomrule
    \end{tabular}
\begin{tablenotes}
Standard errors for proportions in parentheses.
\end{tablenotes}
\end{table}
}

In terms of  $\Sigma$-\emph{dominance}, the lab and Prolific samples are statistically inseparable ($p=0.725$) with approximately one-in-ten participants making a defect choice in the last two games. In contrast, for the MTurk sample this rate grows to more than one-in-three, significantly different from both of the other two populations ($p<0.001$). Moreover, as we explain next, even this number is perhaps an underestimate of the fraction of participants making choices orthogonal to the incentives.

Where the \emph{Average} column in Table \ref{tab:PooledCoop} provides the overall average results by population sample (pooling across both the $C$-first and $D$-first frames), the \emph{Re-framing} column indicates the change in the proportion across the re-frame (the change in the participant proportion exhibiting a $\Sigma$-dominated choice when we move from listing $C$ to listing $D$ as the first action, and the arrows in Figure \ref{fig:CoreChoice}(A)  mirror this number). Our results across the re-frame show that the lab sample moves in the opposite direction from a first-option bias with a slight decrease in  $\Sigma$-dominance when the $D$ action is listed first (though this is not significant, $p=0.614$). The Prolific sample does show a movement 5.2 percentage point movement, where 15.6 percent of choices in the $D$-first sample are $\Sigma$-dominated choices. Though this difference is not significant ($p=0.160$) if we allowed for a one-sided test there is marginal evidence for a small first-action bias on Prolific. The largest effect though is in the MTurk sample, where listing the $D$-action first leads to a 16.3 percentage point increase in the  $\Sigma$-dominated fraction ($p<0.001$ on a test of proportions).\footnote{While we focus more on behavior in each separate game below, the bottom section of Panel B in Table \ref{tab:PooledCoop} indicates that the re-framing has a consistent effect in increasing the selection of $D$ in all four games in the MTurk sample when this action is listed first.}

In the worst-case $D$-first treatment 47.8 percent of the MTurk choices are $\Sigma$-dominated. Despite successfully passing the screen questions---where participants must demonstrate their understanding of the game incentives or be kicked out---approximately one half of the MTurk sample then make choices that indicate little awareness of the induced games. While approximately a third of this effect can be attributed to participants choosing the $D$ action in games \G{3} and \G{4} simply because it is the first-listed option, the result still indicates that just under half  of the sample are making choices that are orthogonal to the offered incentives. In contrast, despite similar costs per observation on Prolific, the rate of such mistakes in this population seems to be at most 15 percent, and we lack statistical power to say that it is even different from the laboratory.

We summarize these first two results:
\begin{result}[$\Sigma$-Dominance] The MTurk sample exhibits significantly more choices that violate any other-regarding preference that seeks to maximize efficiency than the other two populations. The Prolific sample is not significantly different from the Lab sample based on this data-quality measure. 
\end{result}

\begin{result}[Response to frame] The MTurk sample exhibits significantly more choices that select the first-listed option. While there is also a small effect for the Prolific sample, the effect is only marginally significant. We do not detect any effect in the lab sample from the re-framing. 
\end{result}

The focus of the above is on measuring the extent to which choices are being driven by mistakes---essentially any non-designed feature of the strategic environment that is orthogonal to the economic payoff variables. We now examine comparisons that may be more indicative of preference differences across the population. We start by outlining the level effects on two other choice profiles: (i) satisfying $i$-dominance (equivalent to Nash choice in our games); (ii) a partial cooperator who defects in game \G{1} but cooperates in game \G{2} when the PD tensions are smaller, the choice profile that identifies the Rapoport hypothesis; and (iii) those fully cooperating across the four games. We then  summarize the fraction of participants whose actions are (iv) $\Sigma$-dominant and (v) both $\Sigma$-dominant and correctly ordered by the Rapoport measure (the sum across the $DD\mhyphen CC$, $DC\mhyphen CC$ and $CC\mhyphen CC$ profiles). 

In terms of selfish Nash-like behavior, the largest proportion is found in the lab sample. Here 32.4 percent of the lab participants choose the $DDCC$ $i$-Dominant profile. This compares to 26.0 percent in the Prolific sample ($p=0.273$ compared to the lab) and 15.9 percent in MTurk ($p=0.004$).\footnote{If we restrict attention to only those participants satisfying $\Sigma$-dominance, the proportion of $i$-Dominant play in the three samples is much more comparable across the three populations. However, in this sub-sample, the lab proportion of selfish play is still marginally higher ($p=0.078$)}. The rate of fully selfish play is not significantly affected by the re-framing treatment in any of the three populations.

The prediction based on the behavioral literature (primarily lab-based studies) is that large proportion of participants will exhibit a form of partial cooperation, defecting in the first game (Rapoport ratio of 0.5) but cooperating in the other games (Rapoport ratio of 0.71 in game \G{2}). Given this, we label the choice profile $DC\mhyphen CC$ as the Rapoport identifying profile, where the quantitative prediction from the prior literature is that 14 percent of participants were expected to have this exact choice profile. In fact, our lab sample exceeds this level (though we cannot reject it, $p=0.280$). In contrast, both the MTurk and Prolific samples shower a smaller rate on this choice profile, closer to ten percent incidence in each (significantly different from the 14 percent prediction on each). Comparing the three populations, we reject equality of the proportions when we compare the lab to each of the two samples collected online ($p=0.042$ for MTurk and $p=0.083$ for Prolific).\footnote{We do find an effect in the MTurk sample from the re-framing treatment for the Partial Cooperator proportion, decreasing by 5 percentage points when $D$ is listed first ($p=0.049$).
}

Finally in the \emph{Full Cooperator} rows in Table \ref{tab:PooledCoop} we outline the proportion of participants that cooperate in all four games. Here the lab has the smallest fraction of full-cooperators, followed by MTurk ($p=0.807$ in comparison to the lab level), where Prolific has the most ($p=0.024$).\footnote{Looking at the re-framing, while there is a 5 percentage point reduction in fully cooperative profiles, the result is not significant ($p=0.262$). While the MTurk cooperation rate in every single game declines by approximately 14 percentage points under the re-framing, subjects exhibiting this effect are not simply choosing the first-listed action in each game, as otherwise we would see a stronger effect here.}

The final sets of results outline collections of types. The $\Sigma$-dominant row is simply the complement of the first row, outlining the fraction of participants that chose to cooperate in both games \G{3} and \G{4}. In the \emph{Rapoport Ordered} rows we focus on behavior in the PD games (\G{1} and \G{2}), measuring the proportion of participants whose choices are ordered by the Rapoport index (so $CC$, $DC$ or $DD$ in the PD games). Finally, the \emph{Both} row indicates the proportion of participants that satisfy both   $\Sigma$-dominance and the Rapoport ordering (where this is the sum of the three individual choice profiles above). Overall, the conclusions inspecting these three rows are  that the laboratory and Prolific samples are similar in terms of these broad categories. In contrast the MTurk sample exhibits a substantial (and significant, $p<0.001$ all comparisons) drop in all three measures.

\begin{result}[Behavior Comparison] The lab sample replicates the literature finding, with a substantial dropoff in cooperation between games \G{2} and \G{1}, where this pattern is not found in either the Prolific or Mturk data.
\end{result}

\section{Discussion}
While the lab sample does replicates the standard comparative static over the Rapoport ratio, it does so with a relative lack of power due to the more substantial cost per observation. In contrast, both of the online samples show a close-to-zero effect over the PD comparison, despite what should be much greater power even if the effect were substantially attenuated. What then can we conclude? 

To begin we focus on quantifying the pure noise component, the proportion of choices that is made randomly, and entirely orthogonal from the offered incentives. We then separate this component from what is potentially just inelasticity in response, attenuation in the effect size driven by participants in the online populations simply having different preferences. 

To assess the pure noise effects, we return to the behavior in games \G{3} and \G{4} where there is no real strategic tension. We suppose that there are three types. (i) An inattentive type that chooses the first-listed option regardless of the offered incentives, with measure $\gamma_F$ in the population. (ii) an inattentive type that randomly chooses one of the two options in each game regardless of the incentives, measure $\gamma_R$ in the population. (iii) An attentive type that responds to the incentives and satisfies $\Sigma$-dominance, with incidence $\gamma_\Sigma=1-\gamma_F-\gamma_R$.

Using a simple mixture model over these three types, we then estimate the mass on each of the three types using the population samples.\footnote{The structural model thereby accounts for a one-in-four chance that a random type that would be classified as $\Sigma$-dominant.} Our model estimates indicate:
\begin{description}
    \item[Lab] For the lab sample we estimate: $\hat{\gamma}_F=0.000$, $\hat{\gamma}_R=0.144$ and   $\hat{\gamma}_\Sigma=0.856$.
    \item[Mturk] For the MTurk sample we estimate: $\hat{\gamma}_F=0.082$, $\hat{\gamma}_R=0.512$ and   $\hat{\gamma}_\Sigma=0.406$.
    \item[Prolific] For the Prolific sample we estimate: $\hat{\gamma}_F=0.015$, $\hat{\gamma}_R=0.180$ and   $\hat{\gamma}_\Sigma=0.805$.
\end{description}

Thinking of the noise in the data simply due to inattentive participants, we find that MTurk data is almost 60 percent noise, so just 40 percent of the respondents make choices driven by the offered economic incentives. In contrast, the signal component in the data is 81 percent for the Prolific sample and 86 percent for the laboratory sample (in the sense of satisfying $\Sigma$-dominance).\footnote{Bootstraps over the estimation indicate standard errors for the responsive proportion of 4.9 percent in the lab sample, 3.3 percent for MTurk and 2.6 percent for Prolific.} 

While the proportion of random decisions is certainly large in the MTurk sample, each observation is very cheap. At \$22 per lab observation, we can collect seven MTurk observations for every one in the lab. As such, if the minority of the MTurk sample that \emph{is} incentive responsive have a similarly sized response over the two PD games to the prior literature, then the Mturk sample would dominate the lab sample in power terms. At a cost per observation of \$3.01 and an attenuation effect of 59.4 percent the MTurk population should still have 88 percent power for detecting a response when comparing behavior in the PD games, compared to 75.3 percent power for the lab sample (at 14.4 percent noise). Both though would be clearly dominated by observations from Prolific. At  \$4.36 per observation and 19.5 percent noise, if the PD-game effect size on Prolific matched the previous literature our Prolific sample should yield a test with 99.7 percent power.

\ShowIf{
\begin{figure}[tb]
    \centering
    \begin{subfigure}{0.49\textwidth}
        \includegraphics[width=1\textwidth]{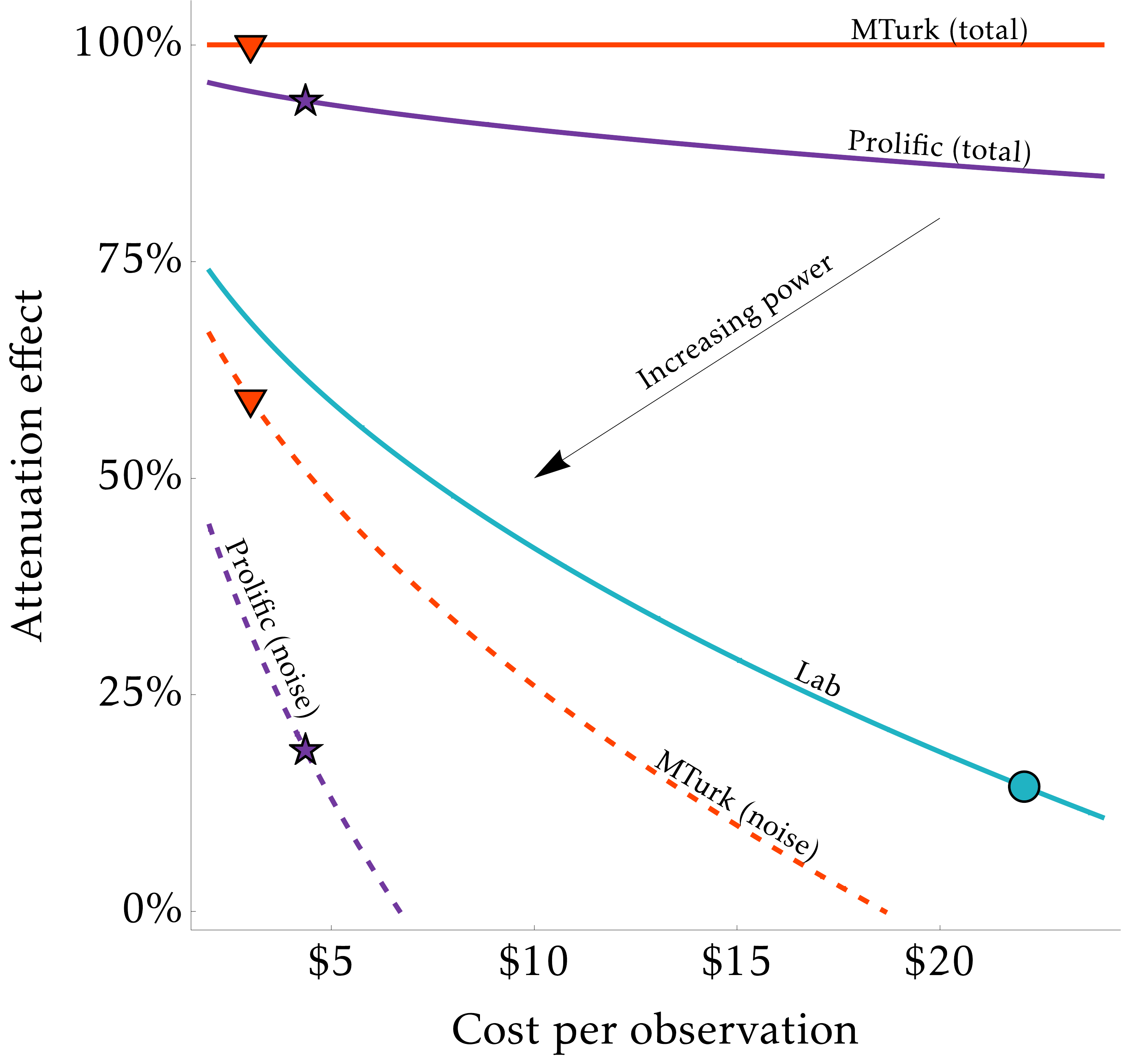}
        \caption{Original comparison}
    \end{subfigure}
    \begin{subfigure}{0.49\textwidth}
       \includegraphics[width=1\textwidth]{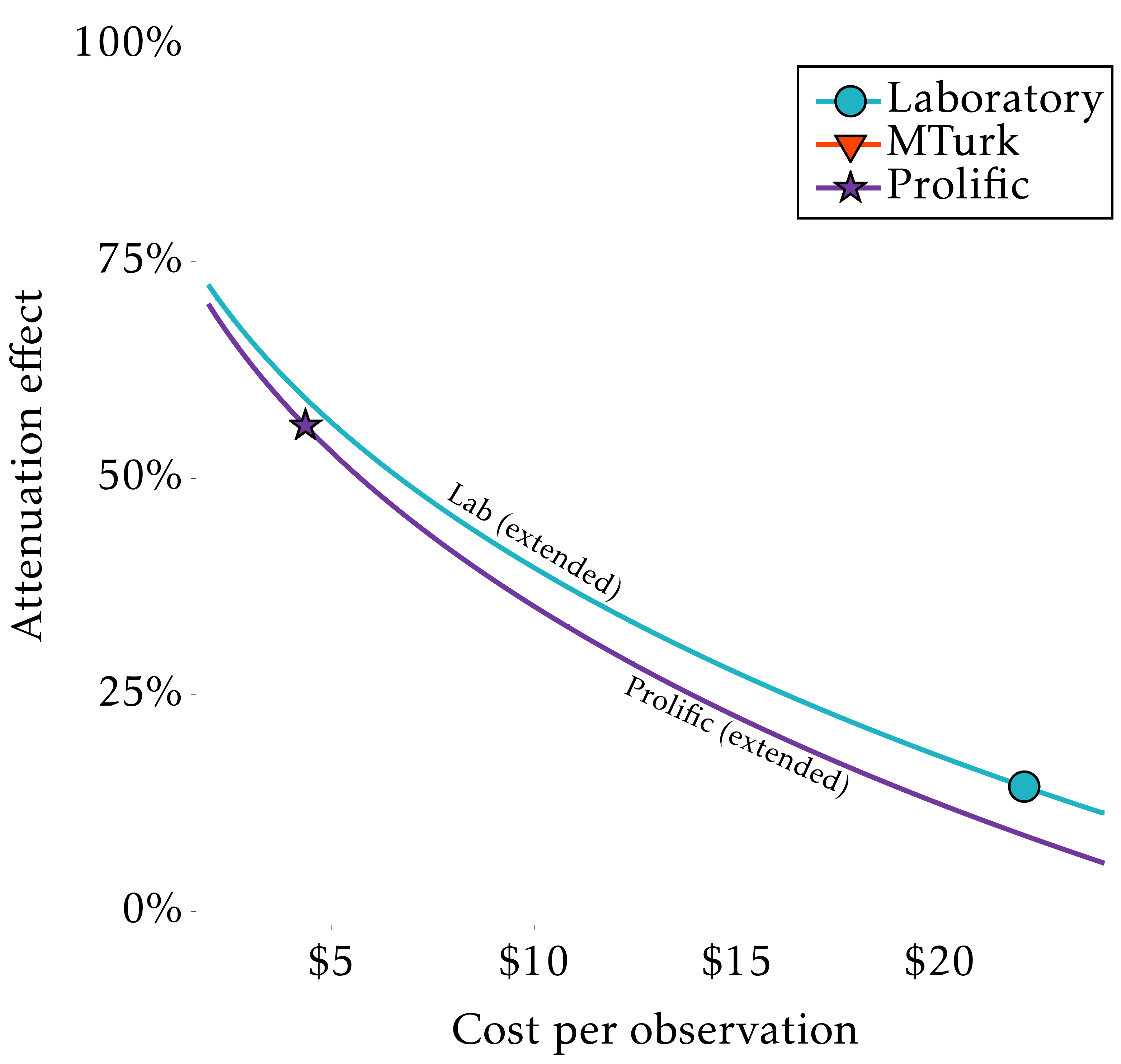}
        \caption{Extended comparison}
    \end{subfigure}
    \caption{Population power\label{fig:PowerActual}}
\begin{tablenotes}
Dotted lines show effective power if the populations only differ by the pure noise effects (and true rates reflect the prior laboratory literature). Solid lines indicate composite attenuation, where inelasticity in response in non-lab samples also factored in. 
\end{tablenotes}
\end{figure}
}

The dotted lines in Figure \ref{fig:PowerActual}(A) represent iso-power countours over populations with $(c,\gamma)$ where we maintain the prior-literature effect sizes as the true effect size. Each line therefore represents the equivalent population to MTurk (orange) and Prolific (purple) samples where the labeled points indicate our estimates for each population. The curve tells us that at 19.5 percent attenuation Prolific would still be preferable to MTurk even if its cost per observation increased to \$12.20. Alternatively, fixing the current MTurk cost the noise would have to shrink to 32.6 percent to match the Prolific sample's power. Both online samples would dominate the laboratory if their true effect sizes  were similar to the prior lab literature, illustrated here by the solid blue line (with the blue circle indicating our estimates for $c$ and $\gamma$). However, noise due to inattention is not the sole factor to consider. Not only do we want a large fraction of attentive participants that are responsive to offered incentives, we also need that population to show an elastic response across our hypothesis.\footnote{For example, see \citet{araujo2016slider} who demonstrate that while a particular real-effort task does show a qualitative response to incentives, the effect sizes are economically too small for it to be an effective tool.} The true attenuation effects within the populations are an amalgam of both the noisy/inattentive participants, and any reduction in the effect size. 

The solid lines in  Figure \ref{fig:PowerActual}(A) represent the iso-power lines when we calculate the total attenuation relative to the lab. Here we calculate the critical attenuation rate $\gamma$ if the true effect matches the lab response that produces the same power test given the realized average response on each of our online populations. This substantially changes the ranking across our populations. MTurk is entirely unresponsive as the realized levels actually have the opposite sign from the hypothesis, and so full attenuation generates the most powerful test. In contrast Prolific does show a small amount of power, but because the difference in cooperation between games \G{1} and \G{2} is just 1 percentage point,\footnote{Exacerbating the effect, cooperation rates are closer to 50 percent, causing the maximal standard error for a proportion test.} the laboratory sample has much greater power. When it comes to making inference over a relatively simple self/other strategic tradeoff, while the Prolific sample does exhibit substantial internal consistency, the quantitative response to the Rapoport ratio is tiny.

One possible conclusion though is simply that online samples do not respond to social-dilemma tensions in the same way as laboratory participants do. As such, the Rapoport ratio finding may be a lab-specific phenomenon. To examine this, and check that there is a response in more extreme games, we ran a second Prolific study. Recruiting a further 125 participants, we added two PD games to the previous four games. In the two additional games we ramp up the PD tensions, so that the Rapoport ratios of the added games are 0.05 and 0.25.\footnote{The precise games are given in the Online Appendix}. Looking to \citet{charness2016social} for the prior literature effect size, we estimate that a comparison of the most extreme PD games (Rapport ratios of 0.71 and 0.05) in a lab sample should show a cooperation reduction of 48 percentage points. 

We do find a more substantial effect in the second Prolific study. In the more-extreme PD games the cooperation rate falls to 0.320, with a comparable cooperation rate 0.584 in the least extreme PD game.\footnote{We can not reject the original cooperation levels over \G{1} and \G{2}, despite the increase to 6 choices.} While the difference in cooperation  is now highly significant ($p<0.001$), the 26 percentage point reduction represent approximately half the effect size we would expect in the lab sample over the same comparison. We illustrate this alternative comparison where we ramp up the power of the test through more-extreme game choices in Figure \ref{fig:PowerActual}(B). 

Modeling the lab as having an attenuation purely driven by noise (0.144) and a cost per observation of \$22, a comparison of the two most-extreme PD games with a true cooperation rate difference of 48 percentage points yields a near certain test when the budget is \$1,650 ($>99.99$ percent power). Despite a reduced effect size on Prolific, the cheaper observations yield almost the same effect. For clarity then we switch to thinking about the dual problem here: what is the experimental budget that yields a 95 percent power test in each population on this more-extreme comparison? A budget of \$379 on Prolific yields the same 95 percent power test as a \$484 budget in the laboratory. This slight gain for Prolific over the lab is illustrated by the lower curve in Figure \ref{fig:PowerActual}(B). 

Our results on the extended games suggest that online populations \emph{are} capable of uncovering the same qualitative patterns as the laboratory. However, two caveats are appropriate here. First, the substantial noise on MTurk suggests that more-curated populations are likely superior (or that greater internal validity checks are required, where sub-sampling the population substantially increases the cost of each valid MTurk observation). Second, the elasticity of response to other-regarding tensions in these online populations is much-attenuated from previous lab samples. Under more-nuanced parameterizations, the online populations' response is simply too small to have power, where the lab sample shows much greater response. If the aim is purely to uncover qualitative findings, or to gauge the order of magnitude of an effect, the conclusion from our study is to eschew all subtlety. So long as the parameterization can generate a moderate effect size, much greater power can be produced by the smaller cost per observation.

On the flip-side of the coin, our study also points to the usefulness of the laboratory. Certainly in our setting, lab participants tend to have stronger responses. In studies where the aim is to educe more nuanced findings---calibrating a non-linear model say, where estimating curvature requires smaller step-size in the treatments---the lab can play a useful role. Despite the expense per observation, the combination of more-elastic response to shifts in the incentives and a low noise rates make the lab a better tool. While our study has no variation in the level of complexity, the lab offers a conducive environment for testing knottier economic hypotheses. By controlling participants' outside-option activities and removing distraction, lab samples allow for experimenters to induce more-complex artificial settings. While there is certainly a place for online samples, given their low cost and ease of acquisition, a lack of control on attention does seem to be a problem.

\section{Conclusions}\label{sec:conc}
We examine three populations for conducting experimental studies. Rather than a validation of comparative statics across the differing populations, we take a different tack. We reflect ecological differences in the price for each observation and the noise in samples from each by constructing the researcher's preference over populations via inferential power. That is, to what extent might one want to trade off some level of noise for much cheaper observations, assuming the researcher has a fixed budget to spend.

Our design measures both the noise in each population (which we attribute to inattention through a weak assumption on preferences) and a more-nuanced hypothesis on the response to social dilemma tensions. We fix the experimental budget on each population and, by varying the scale of the incentives so that they match standard levels in each population, we thereby vary the number of observations collected on each. We then assess the noisiness in behavior and the extent to which each population replicates a standard comparative static results from the literature. 

Our findings indicate that a small number of participants in a laboratory sample (with relatively expensive observation costs) replicate standard findings. However, two online samples with lower observation costs (MTurk and Prolific) do not. In terms of the proportion of noise in the data, both the laboratory and Prolific have relatively low levels (14 and 20 percent, respectively). In contrast, at 60 percent our MTurk sample is particularly noisy, despite screens to ensure understanding. However, even at this level of noise, the very cheap observations from MTurk should dominate the laboratory from an inference point of view---though in pure terms of noise versus cost, Prolific dominates both.

However, beyond just noise, our lab sample exhibits much greater elasticity of response to treatment where the online samples are essentially inelastic on that margin. While these results may be specific to social dilemmas---where our more-generalizable noise estimate clearly outline the power benefits of Prolific samples---they outline that despite the expense per observation, lab samples can offer greater power. However, we go on to show that by making the size of the treatment effect larger Prolific can again dominate the lab in power terms, despite the reduced elasticity of response. 

The very substantial noise in the MTurk sample may be a recent phenomenon (where studies suggest the population has recently declined). However, our analysis suggests that despite being the cheapest of the samples, MTurk offers a false economy. While almost 50 percent more expensive, Prolific observations offers substantially greater inferential power by reducing noise. While our study indicates there are domains where lab data is preferable, if an online population is to be sampled, our study clearly indicates that more-curated populations such as Prolific are superior to MTurk.


\ShowIf{
\bibliographystyle{aer}
\bibliography{./lit/lit}

\appendix
\let\oldthetable\thetable
\let\oldthefigure\thefigure
\setcounter{table}{0}
\setcounter{figure}{0}
\renewcommand{\thetable}{A.\oldthetable}
\renewcommand{\thefigure}{A.\oldthefigure}
\vfill\eject
\clearpage
\section{Additional Figures and Tables}

\begin{figure}[htb]
    \centering
    \label{fig:ChoiceProfile}
    \includegraphics[width=0.95\textwidth]{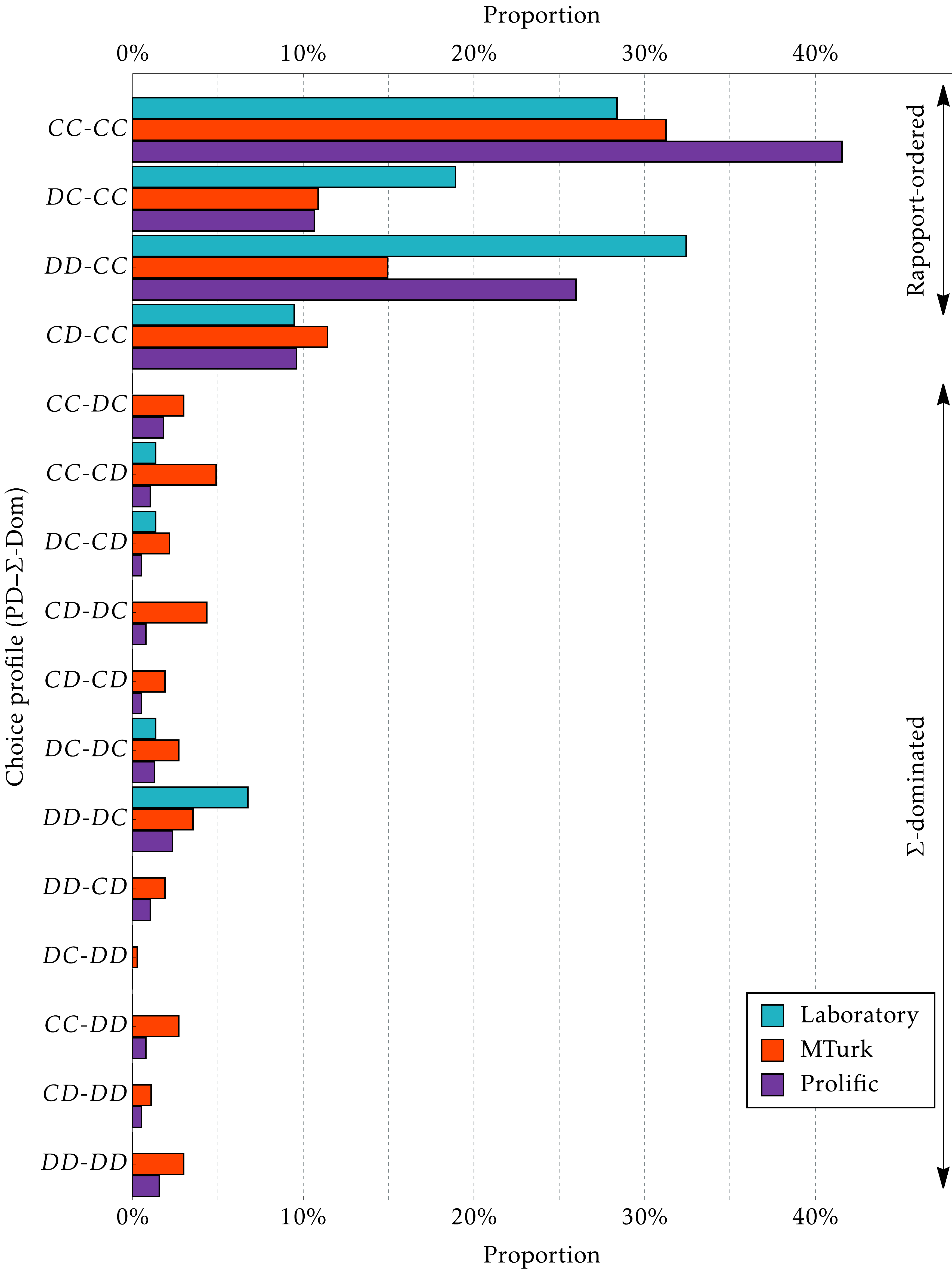}
    \caption{Choice Profiles: By Population}

\end{figure}

\subsection{Prolific Robustness Sessions: Extended Response}\label{app:robust}

\begin{table}[h!]
\centering
\caption{Experimental Games: Robustness Sample}
\label{tab:games_robust}
\begin{tabular}{c|c|ccc|c|c}
\multicolumn{3}{c}{\G{1} game:} &  & \multicolumn{3}{c}{\G{2} game:}\\ 
\multicolumn{1}{c}{} & \multicolumn{1}{c}{\emph{C}} & \emph{D} &  & \multicolumn{1}{c}{} & \multicolumn{1}{c}{\emph{C}} & \emph{D}\\ 
\cline{2-3} \cline{3-3} \cline{6-7} \cline{7-7} 
\emph{C} & 21,21 & \multicolumn{1}{c|}{2,28} &  & \emph{C} & 19,19 & \multicolumn{1}{c|}{8,22}\\ 
\cline{2-3} \cline{3-3} \cline{6-7} \cline{7-7} 
\emph{D} & 28,2 & \multicolumn{1}{c|}{8,8} &  & \emph{D} & 22,8 & \multicolumn{1}{c|}{9,9}\\ 
\cline{2-3} \cline{3-3} \cline{6-7} \cline{7-7} 
\multicolumn{1}{c}{} & \multicolumn{1}{c}{} &  &  & \multicolumn{1}{c}{} & \multicolumn{1}{c}{} & \\ 
\multicolumn{3}{c}{\GR{5} game:} &  & \multicolumn{3}{c}{\GR{6} game:}\\ 
\multicolumn{1}{c}{} & \multicolumn{1}{c}{\emph{C}} & \emph{D} &  & \multicolumn{1}{c}{} & \multicolumn{1}{c}{\emph{C}} & \emph{D}\\ 
\cline{2-3} \cline{3-3} \cline{6-7} \cline{7-7} 
\emph{C} & 14,14 & \multicolumn{1}{c|}{5,25} &  & \emph{C} & 18,18 & \multicolumn{1}{c|}{3,27}\\ 
\cline{2-3} \cline{3-3} \cline{6-7} \cline{7-7} 
\emph{D} & 25,5 & \multicolumn{1}{c|}{13,13} &  & \emph{D} & 27,3 & \multicolumn{1}{c|}{12,12}\\ 
\cline{2-3} \cline{3-3} \cline{6-7} \cline{7-7} 
\multicolumn{1}{c}{} & \multicolumn{1}{c}{} &  &  & \multicolumn{1}{c}{} & \multicolumn{1}{c}{} & \\ 
\multicolumn{3}{c}{\G{3} game:} &  & \multicolumn{3}{c}{\G{4} game}\\ 
\multicolumn{1}{c}{} & \multicolumn{1}{c}{\emph{C}} & \emph{D} &  & \multicolumn{1}{c}{} & \multicolumn{1}{c}{\emph{C}} & \emph{D}\\ 
\cline{2-3} \cline{3-3} \cline{6-7} \cline{7-7} 
\emph{C} & 17,17 & \multicolumn{1}{c|}{12,16} &  & \emph{C} & 15,15 & \multicolumn{1}{c|}{16,10}\\ 
\cline{2-3} \cline{3-3} \cline{6-7} \cline{7-7} 
\emph{D} & 16,12 & \multicolumn{1}{c|}{10,10} &  & \emph{D} & 10,16 & \multicolumn{1}{c|}{11,11}\\ 
\cline{2-3} \cline{3-3} \cline{6-7} \cline{7-7} 

\end{tabular}
\end{table}

\begin{table}[h!]
    \centering
    \caption{Prolific Participants per treatment}
    \label{tab:Subjects_robust}
    \begin{tabular}{cccc}\toprule
         & Prolific & Prolific-Robustness\\ \midrule
        Main    & 250 & 125\\
        Re-frame & 135 & &\\ \bottomrule
    \end{tabular}
    \begin{tablenotes}
Excludes participants who did not answer the comprehension question correctly. \end{tablenotes}
\end{table}

\begin{table}[h!]
\centering
\caption{Behavior Across Prolific Samples: Cooperation }
\label{tab:PooledCoop_robust}
\begin{tabular}{lccc}
        \toprule  Game & Robustness Sample & p-value & Original Sample\\
        \midrule
        \G{1} & $\mEst{0.488}{0.045}$ & 0.127 & $\mEst{0.566}{0.025}$ \\
        
        \G{2} & $\mEst{0.584}{0.044}$ & 0.885 & $\mEst{0.577}{0.025}$ \\
        
        \G{3} & $\mEst{0.904}{0.026}$ & 0.865 & $\mEst{0.909}{0.015}$ \\
        
        \G{4} & $\mEst{0.952}{0.019}$ & 0.623 & $\mEst{0.94}{0.012}$ \\
        \midrule
        \GR{5} & $\mEst{0.320}{0.042}$ &  &  \\
        
        \GR{6} & $\mEst{0.328}{0.042}$ &  &  \\
        \bottomrule
    \end{tabular}
\begin{tablenotes}
Standard error for the proportion in parentheses. All $p$ values are for two-sided tests of equality between the samples. \end{tablenotes}
\end{table}

\clearpage
\begin{table}[h!]
    \centering
    \caption{Subject Types Across Prolific Samples: Pooled Data}
    \label{tab:SubjectTypes_robust}
    \begin{tabular}{lccc}
        \toprule  Type & Robustness Sample & p-value & Original Sample\\
        \midrule
        \multicolumn{4}{l}{Choice Profiles in Original 4 Games:}\\
        \quad\quad Nash $(DD\mhyphen CC)$ & $\mEst{0.312}{0.042}$ & 0.255 & $\mEst{0.260}{0.022}$ \\
        
        \quad\quad Uncond Coop ($CC\mhyphen CC$) & $\mEst{0.352}{0.043}$ & 0.208 & $\mEst{0.416}{0.025}$ \\
        
        \quad\quad Cond Coop ($DC\mhyphen CC$ \& $CD\mhyphen CC$) & $\mEst{0.208}{0.036}$ & 0.897 & $\mEst{0.203}{0.021}$ \\ 
        
        \quad\quad $\Sigma$-dominated & $\mEst{0.128}{0.030}$ & 0.862 & $\mEst{0.122}{0.017}$ \\
        \bottomrule
 \end{tabular}
\begin{tablenotes}
Standard error for the proportion in parentheses. All $p$ values are for two-sided tests of equality between the populations. Choice profiles ar given in order of the Rapoport ratio in the PD games (so \GR{5}, \GR{6}, \G{1}, \G{2}-\G{3}, \G{4}).
\end{tablenotes}
\end{table}

\begin{table}[h!]
    \centering
    \caption{Additional Subject Types in Prolific Robustness Sample}
    \label{tab:SubjectTypes_robust_additional2}
    \begin{tabular}{lc}
        \toprule  Type & Robustness Sample\\
        \midrule
        $\Sigma$-dominant & $\mEst{0.872}{0.030}$ \\ 
       Rapoport ordered & $\mEst{0.808}{0.035}$\\
       Both & $\mEst{0.728}{0.040}$\\
        \multicolumn{2}{l}{$\Sigma$-dominant profiles:}\\
        \quad\quad Nash, $DDDD$ & $\mEst{0.272}{0.040}$  \\
        \quad\quad $DDDC$ & $\mEst{0.120}{0.029}$  \\
        \quad\quad $DDCC$ & $\mEst{0.088}{0.025}$  \\
        \quad\quad $DCCC$ & $\mEst{0.072}{0.023}$  \\
        \quad\quad Uncond Coop, $CCCC$ & $\mEst{0.176}{0.034}$ \\
        \quad Non-Rapoport ordered (11 profiles) & $\mEst{0.137}{0.030}$\\
        \bottomrule
 \end{tabular}
\begin{tablenotes}
Standard error for the proportion in parentheses. Choice profiles ar given in order of the Rapoport ratio in the PD games (so \GR{5}, \GR{6}, \G{1}, \G{2}-\G{3}, \G{4}).
\end{tablenotes}
\end{table}



\clearpage
\section{Instructions for Laboratory Experiment} \label{app:lab}

\subsection{Instructions for Main Treatment}\label{app:lab_main}
\paragraph{Welcome and thank you for participating in this study. This is an experiment on decision making. Please turn off your cell phones and similar devices now and place them on the top shelf of your station. Please do not talk to or in any way try to communicate with other participants in the room.}

Your earnings in today’s experiment will depend on your decisions, the decisions of others in the room, and on chance. Any money you make will be paid privately and in cash at the end of the experiment. We will start with a brief description of your task today. If you have any questions, please raise your hand and we will come to answer you in private.

\textbf{Explanation of your task}\\
There are four rounds in today’s study, each consisting of a decision table. Your task will be to choose one option from two alternatives for each decision table. A round will end when all participants submit their choices.

At the end of the fourth round, the computer will randomly and anonymously pair you with another participant in the room. Next, the computer will randomly select one of your four rounds. You will be paid for that round based on you and the matched participant’s choices in that round. Your final earnings will then consist of payoff from this one round and a participation fee of \$6. 

Every round is equally likely to be selected for payment, so you should treat each round as if it determines your final payment. Also, there are only four decisions in this study, so you should consider them carefully.

\textbf{Description of a Decision Table}\\
Below is an example decision table:
\begin{figure}[h!]
    \centering
    \includegraphics[scale=1.25]{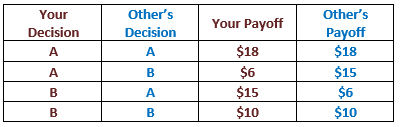}
\end{figure}
Both you and the matched participant make choices between Option A and Option B. The decision table indicates the payout for you and the other participant for each possible combination of choices.

Suppose this decision table was selected for payment, then in addition to the participation fee:
\begin{itemize}
    \item[(1)] if both participants choose A, they each receive \$18;
    \item[(2)]	if you choose A and the matched participant chooses B, then you receive \$6, and they receive \$15;
    \item[(3)]	Vice versa if you choose B and the matched participant chooses A, then you receive \$15, and they receive \$6.
    \item[(4)]	if both participants choose B, they each receive \$10;
\end{itemize}
We will begin the study with a few questions about your understanding of the decision table and then proceed to the first round.

\subsection{Instructions for Re-framed Treatment}
[Introductory instructions and section with ``Explanation of your task" were identical to \ref{app:lab_main} ]

\textbf{Description of a Decision Table}\\
Below is an example decision table:
\begin{figure}[h!]
    \centering
    \includegraphics[scale=1.25]{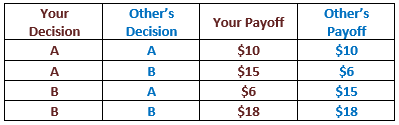}
\end{figure}
Both you and the matched participant make choices between Option A and Option B. The decision table indicates the payout for you and the other participant for each possible combination of choices.

Suppose this decision table was selected for payment, then in addition to the participation fee:
\begin{itemize}
    \item[(1)] if both participants choose A, they each receive \$10;
    \item[(2)]	if you choose A and the matched participant chooses B, then you receive \$15, and they receive \$6;
    \item[(3)]	Vice versa if you choose B and the matched participant chooses A, then you receive \$6, and they receive \$15.
    \item[(4)]	if both participants choose B, they each receive \$18;
\end{itemize}
We will begin the study with a few questions about your understanding of the decision table and then proceed to the first round.

\clearpage
\subsection{Screenshots of the Laboratory Experiment}
Following are the screenshots of the lab experiment for the main sample. The screens for the re-framed sample were identical except that the labels of options on the decision table reversed.\\

\begin{figure}[h!]
    \centering
    \frame{\includegraphics[scale=0.75]{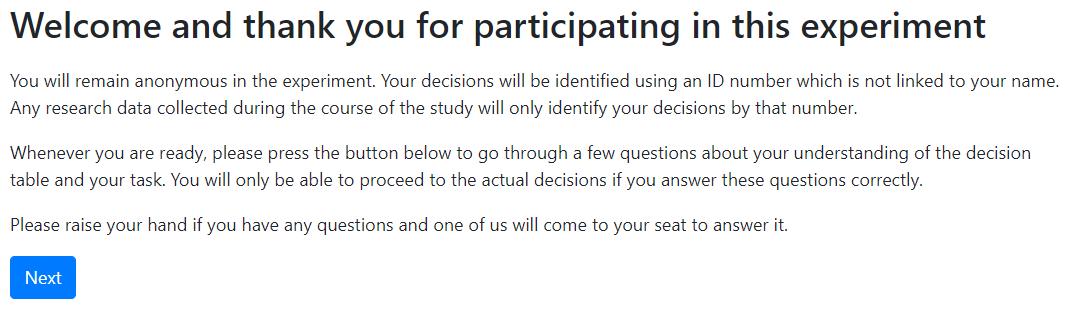}}
\end{figure}

\begin{figure}[h!]
    \centering
    \frame{\includegraphics[scale=0.80]{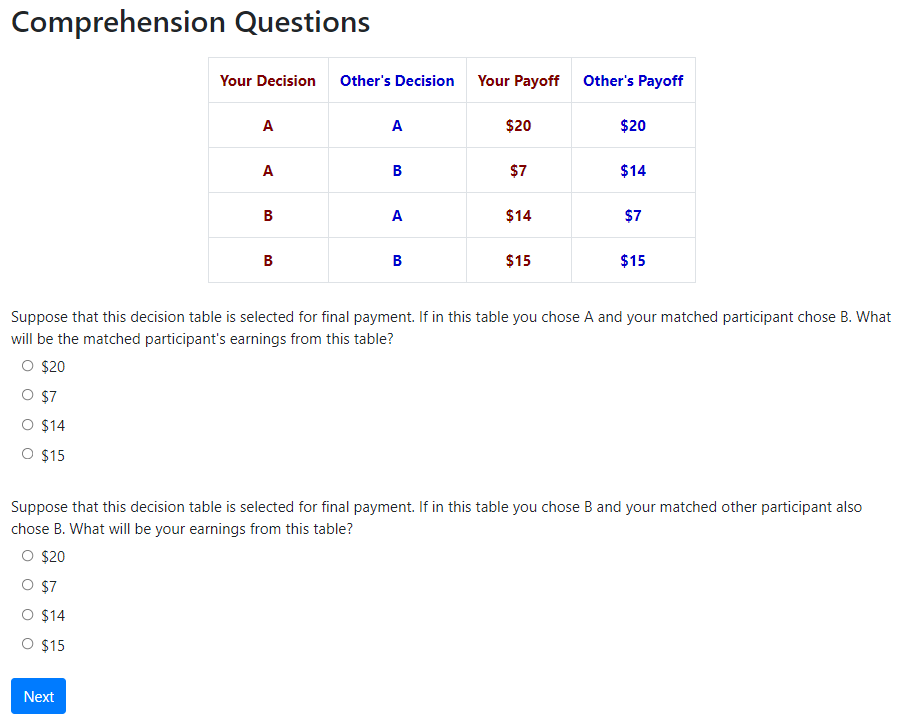}}
\end{figure}
[For the re-framed sample option $A$ corresponded to $D$ and option $B$ corresponded to $C$. The answers to the comprehension questions changed accordingly. Participants couldn't move forward without answering these questions correctly.]

\begin{figure}[h!]
    \centering
    \frame{\includegraphics[scale=0.75]{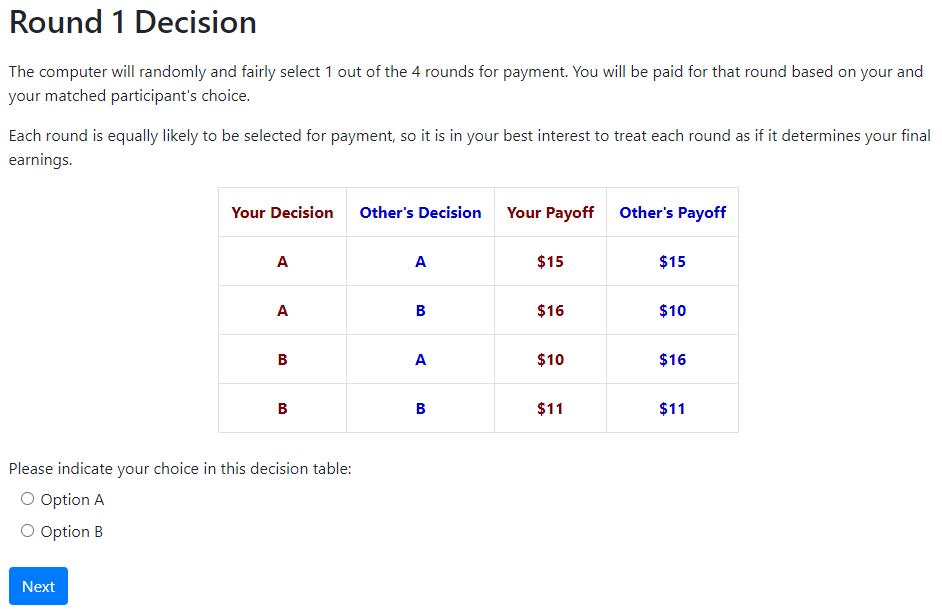}}
\end{figure}
[Rounds 2, 3 and 4 screens were the same as round 1 with different decision tables. For the re-framed sample option $A$ corresponded to $D$ and option $B$ corresponded to $C$, the screens were otherwise the same as the main sample. The four decision tables were presented to the participants in random order.]

\begin{figure}[h!]
    \centering
    \frame{\includegraphics[scale=0.65]{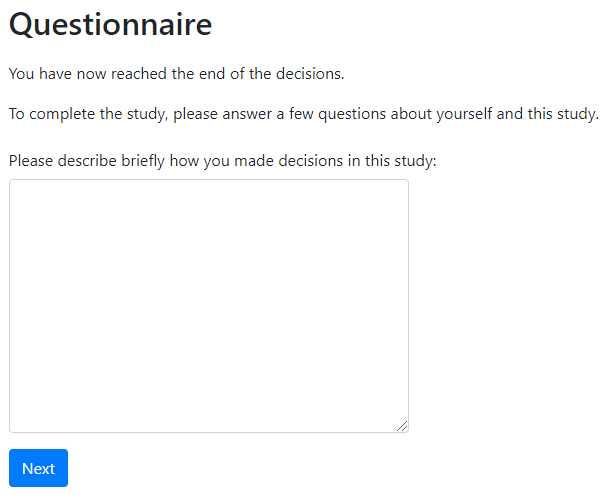}}
    \frame{\includegraphics[scale=0.65]{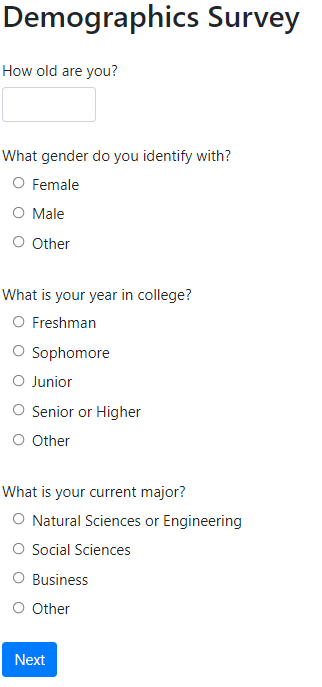}}
\end{figure}

\begin{figure}[h!]
    \centering
    \frame{\includegraphics[scale=0.75]{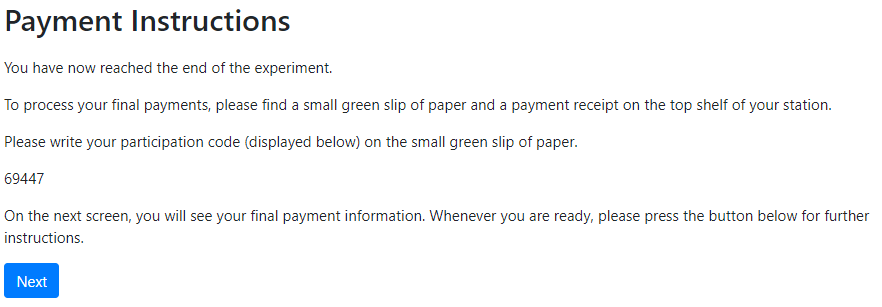}}
\end{figure}

\begin{figure}[h!]
    \centering
    \frame{\includegraphics[scale=0.78]{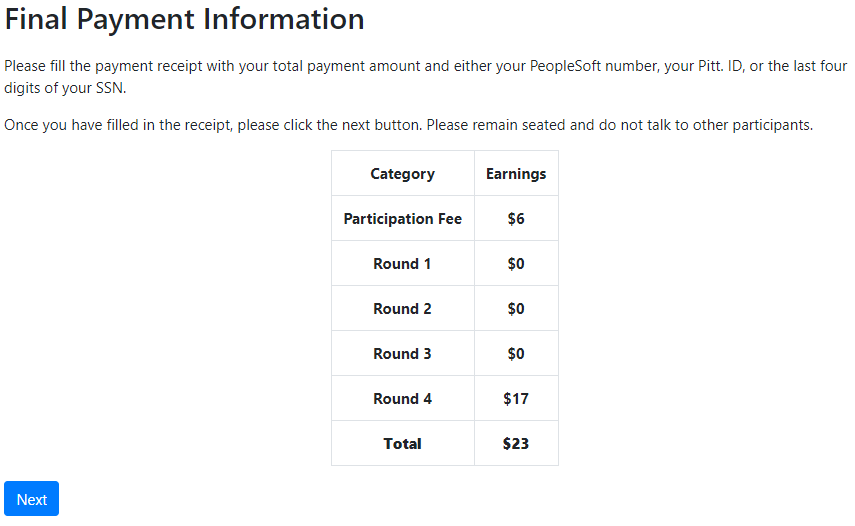}}
\end{figure}
[Participants were then invited to the payment room one by one and paid in cash in private.]

\clearpage
\section{Instructions for Online Experiment} \label{app:online}
Following are the screenshots of the online experiment for the main Prolific sample. The screens for the MechTurk sample were the same as the Prolific sample.\\
\begin{figure}[h!]
    \centering
    \frame{\includegraphics[scale=0.8]{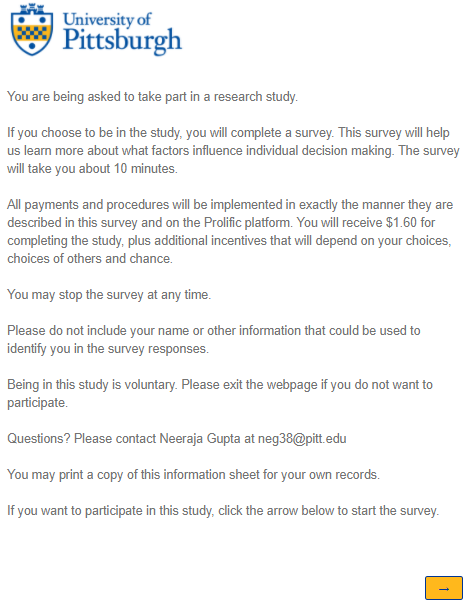}}
    \frame{\includegraphics[scale=0.8]{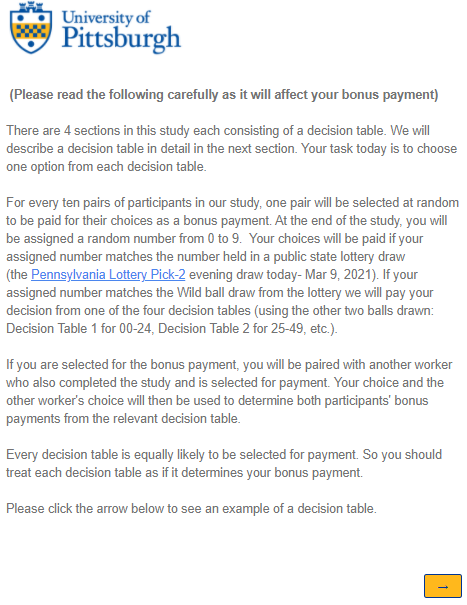}}
\end{figure}

\clearpage
[For the re-framed sample option $A$ corresponded to $D$ and option $B$ corresponded to $C$. The answers to the comprehension questions changed accordingly. Participants were dismissed with the show-up of \$1.60 for answering the comprehension question incorrectly on Prolific (\$0.50 on MechTurk). On MechTurk, participants who answered the comprehension question correctly were offered additional \$0.50.]
\begin{figure}[h!]
    \centering
    \frame{\includegraphics[scale=0.8]{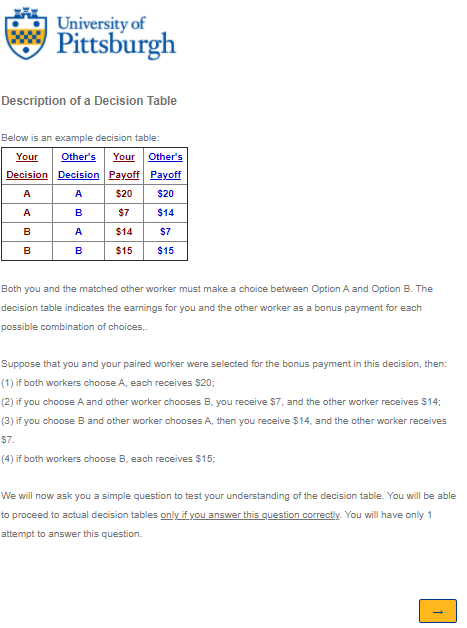}}
    \frame{\includegraphics[scale=0.8]{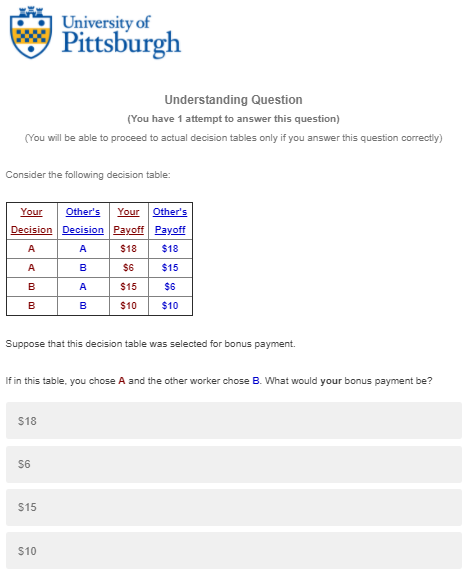}}
\end{figure}

[Next, the four decision tables were presented to the participants in random order. For the re-framed sample option $A$ corresponded to $D$ and option $B$ corresponded to $C$, the screens were otherwise the same as the main sample.]
\begin{figure}[h!]
    \centering
    \frame{\includegraphics[scale=0.8]{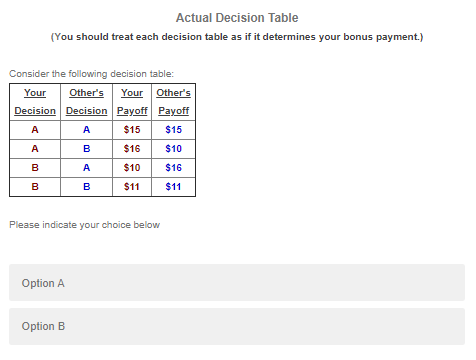}}
\end{figure}

\begin{figure}[h!]
    \centering
    \frame{\includegraphics[scale=0.8]{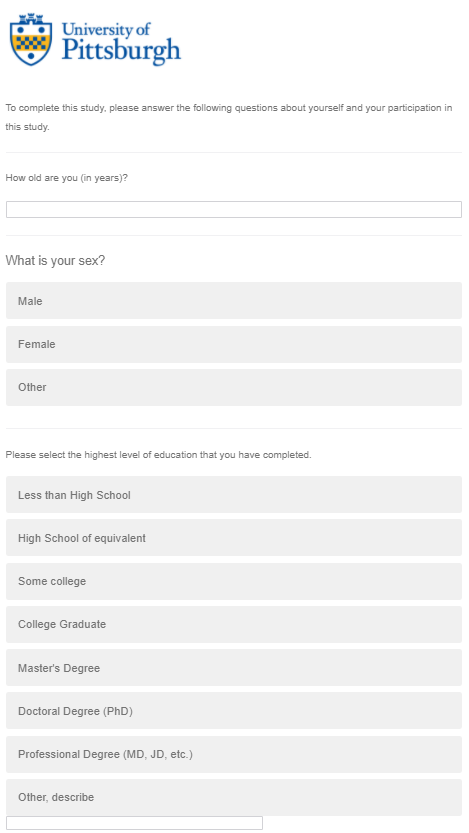}}
    \frame{\includegraphics[scale=0.8]{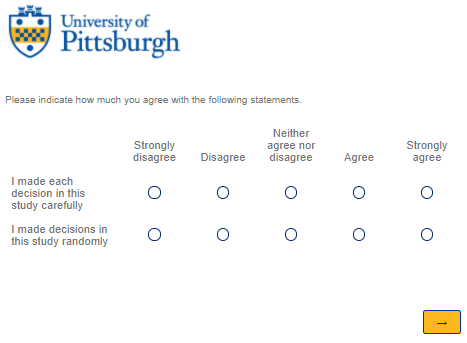}}
    \frame{\includegraphics[scale=0.8]{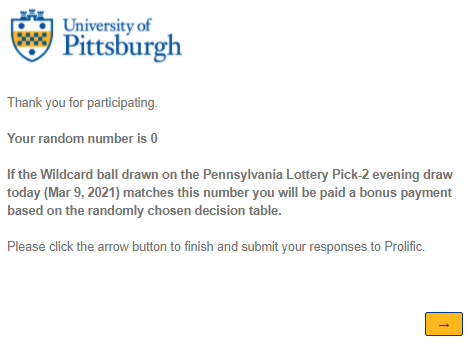}}
\end{figure}
[Fixed fees were credited to the participants immediately upon approval of the submission and the bonus payments were made within 24 hours of completion.]

}
\end{document}